\documentclass[final]{kjmath}

\usepackage{amsmath}
\usepackage{amssymb}
\usepackage{mathtools}
\usepackage{bm}
\usepackage{eucal}
\usepackage{graphicx}
\usepackage{xcolor}
\usepackage{color}
\usepackage{subfig}
\usepackage{float}
\usepackage{placeins}
\usepackage{booktabs}
\usepackage{enumitem}
\usepackage{url}

\newtheorem{assumption}{Assumption}[section]

\newcommand{\R}{\mathbb{R}}
\newcommand{\dd}{\mathrm{d}}
\newcommand{\ain}{\mathfrak{a}}

\newcommand{\certfig}[1]{#1}

\begin{document}

%%%%% DO NOT CHANGE THE FOLLOWING LINES %%%%%%%%%%%
%\kjmvolume{%
%@VOL@
%}
%\kjmnumber{%
%@NUM@
%}
%\kjmyear{%
%@YEAR@
%}
%\kjmstpage{%
%@STPAGE@
%}
%\kjmendpage{%
%@ENDPAGE@
%}
%%%%%%%%%%%%%%%%%%%%%%%%%%%%%%%%%%%%%%%%%%%%%%%%%%%

\title[A novel derivative-corrected within-host viral model]{A novel derivative-corrected within-host\\ viral model based on Caputo--Fabrizio\\ fractional dynamics}

\author[D.\ Sele\v si]{Dora Sele\v si$^{1}$}
\thanks{* Corresponding author.}
\author[D.\ Rajter-\'Ciri\'c]{Danijela Rajter-\'Ciri\'c$^{1}$}
\address{$^1$Faculty of Sciences,
\newline \indent University of Novi Sad,
\newline \indent Serbia}
\email{dora@dmi.uns.ac.rs
	\newline \indent ORCID iD: \url{https://orcid.org/0000-0003-2261-0590}}
\email{rajter@dmi.uns.ac.rs
	\newline \indent ORCID iD: \url{https://orcid.org/0000-0002-2548-8642}}

\author[N.\ Juh\'asz]{N\'ora Juh\'asz$^{2,3}$}
\author[G.\ R\"ost]{Gergely R\"ost$^{2,3}$}
\address{$^2$Bolyai Institute,
\newline \indent University of Szeged,
\newline \indent Hungary}
\address{$^3$Hungarian Centre of Excellence for Molecular Medicine (HCEMM),
\newline \indent Szeged,
\newline \indent Hungary}
\email{juhaszn@math.u-szeged.hu
	\newline \indent ORCID iD: \url{https://orcid.org/0000-0001-6560-7581}}
\email{rost@math.u-szeged.hu
	\newline \indent ORCID iD: \url{https://orcid.org/0000-0001-9476-3284}}

\keywords{Within-host viral dynamics, Caputo--Fabrizio derivative, implicit ordinary differential equations, fractional differential equations, basic reproduction number, peak dynamics, final size relation, long-time behaviour, parameter sensitivity.
%\indent DOI \\
%\indent {\it Received}: \\
%\indent {\it Accepted}:
}

\sloppy

\begin{abstract}
We study an implicit within-host viral dynamics model for uninfected target cells $H(t)$, infected cells $I(t)$, viral load $V(t)$, and damaged or removed cells $D(t)$. The model contains a parameter $\ain\in[0,1]$ that blends the classical instantaneous within-host dynamics with derivative-dependent correction terms. We rewrite the system as an explicit ordinary differential equation whenever a computable determinant $\Delta(H,V)$ is nonzero, derive sufficient conditions for local and global well-posedness, prove positivity and conservation of $H+I+D$, compute the basic reproduction number by the next-generation approach, and analyze long-time behavior. Under natural nondegeneracy assumptions, $I(t)\to0$, $V(t)\to0$, $H(t)\to H_\infty>0$, and $D(t)\to D_\infty$. We also derive peak conditions, a final-size identity, and discuss the biological interpretation of $\ain$ as a dynamical responsiveness or short-memory parameter.
\end{abstract}

\maketitle

\section{Introduction}

Mathematical models have become indispensable tools for understanding the mechanisms governing infectious diseases at different spatial and temporal scales. Depending on the underlying biological question, infectious disease models can broadly be divided into two complementary classes. The first class consists of population-level epidemiological models, whose objective is to describe disease transmission among individuals within a population. These include classical compartmental models such as the susceptible--infected--recovered (SIR) and susceptible--exposed--infected--recovered (SEIR) families \cite{Hethcote2000}, network-based approaches \cite{KeelingEames2005}, agent-based simulations \cite{Eubank2004}, and, more recently, physics-informed neural networks (PINNs) \cite{MuthupandiGnanaprakasamBoulaaras2026}. Such models have been extensively investigated during the past decades and provide quantitative information about epidemic outbreaks, intervention strategies, vaccination policies and disease control.

The second class comprises within-host models, whose purpose is to describe the progression of infection inside a single infected individual. Instead of modelling contacts between individuals, these models investigate the interaction between uninfected (healthy) target cells, infected cells, free viral particles and components of the immune response. Since the pioneering works of Perelson, Nowak, and others \cite{NowakMay2000,PerelsonNelson1999}, and the subsequent development of virus-dynamics models for HIV, influenza, SARS, within-host dynamics models have proved essential for understanding the replication mechanisms of numerous viruses, including also hepatitis B and C, Ebola, dengue, and, since 2020, SARS-CoV-2 \cite{Baccam2006,Beauchemin2011,Ciupe2007,%Perelson2013,
	Rost2021,Yamaguchi2024,Wacker2024}. These models provide quantitative insight into viral replication, disease progression, tissue damage, antiviral treatment and immune response, while simultaneously offering a framework for parameter estimation from experimental viral-load measurements.

Most existing within-host models are formulated as systems of ordinary differential equations, assuming that the rate of change of each biological compartment depends exclusively on the instantaneous state of the system. Such formulations have led to a rich mathematical theory including positivity and boundedness of solutions, threshold analysis via the basic reproduction number, persistence, stability of equilibria, bifurcation analysis, optimal control and parameter estimation. Comprehensive surveys of deterministic within-host viral dynamics may be found in the monograph of Nowak and May \cite{NowakMay2000}, the review by Perelson and Nelson \cite{PerelsonNelson1999}, Wodarz's monograph \cite{Wodarz2007}, and numerous subsequent review papers devoted to mathematical immunology and within-host virus dynamics \cite{Beauchemin2011,CiupeHeffernan2017,EftimieGillardCantrell2016,ZitzmannKaderali2018}.

Although classical ordinary differential equations have proved remarkably successful, increasing biological evidence suggests that many intracellular processes possess memory and hereditary characteristics. Viral replication, intracellular maturation, delayed activation of immune pathways, cytokine signalling and tissue regeneration are influenced not only by the current viral concentration but also by the recent history of the infection. Such phenomena naturally motivate mathematical models capable of incorporating memory into the governing dynamics. Delay differential equations offer a standard approach whenever the exact mechanism of such memory effects is well understood, however such a detailed biological understanding is often lacking. 

Fractional differential equations provide one of the most successful mathematical frameworks for modelling systems with memory. Unlike classical derivatives, fractional derivatives are nonlocal operators depending on the past evolution of the solution, thereby allowing hereditary effects to be incorporated directly into the mathematical model. During the last two decades, fractional differential equations have found widespread applications in viscoelasticity, anomalous diffusion, transport phenomena, control theory, biology, medicine and engineering. Their application to epidemiological and within-host viral models has also expanded rapidly, leading to numerous fractional SIR-, SEIR- and other compartmental epidemic models exhibiting improved agreement with experimental observations in situations where memory effects cannot be neglected \cite{%Diethelm2013Dengue,AngstmannHenryMcGann2016Infectivity,AngstmannEricksonHenryMcGannMurrayNichols2017,
	Chen2021}.

Numerous definitions of fractional derivatives have been proposed in the literature, including the classical Riemann--Liouville and Caputo derivatives \cite{%SamkoKilbasMarichev1993,
	Kilbas2006,Podlubny1999}, the modern generalized derivatives by Luchko \cite{Luchko2020GeneralDerivative}, as well as more recent operators possessing non-singular kernels, such as the Caputo--Fabrizio %and Atangana--Baleanu 
derivative \cite{CaputoFabrizio2015,LosadaNieto2015%AtanganaBaleanu2016
}. While classical fractional operators involve singular power-law kernels corresponding to long-range memory, non-singular fractional derivatives describe fading-memory processes whose influence decreases continuously with time. In many biological applications, such behaviour appears more realistic, since recent events generally exert a stronger influence on the present state than events occurring far in the past, similarly to fading memory in delay differential equations \cite{Gergely}. Among these operators, the Caputo--Fabrizio fractional derivative has received considerable attention owing to its exponential non-singular kernel, analytical tractability and successful applications in diffusion theory, heat transfer, viscoelasticity, pharmacokinetics, biological systems and epidemiological modelling \cite{AlzahraniElDessokyBaleanu2021,AtanackovicJanevPilipovic2019,CaputoFabrizio2016Applications,%MirzaVieru2017AdvectionDiffusion,,%HarroucheMomaniHasanAlSmadi2021,%BaleanuJajarmiMohammadiRezapour2020,
	Moore2019%,ShafqatAlQuranAlsaadiDjaouti2026
}. In the context of within-host viral dynamics, the Caputo--Fabrizio operator offers an attractive framework for describing finite-memory effects associated with viral replication, intracellular development and delayed cellular responses while avoiding the singular kernel characteristic of classical fractional derivatives.

This paper suggests a simple within-host viral infection model formulated by means of the Caputo--Fabrizio fractional derivative. We have announced this model in a short technical report \cite{Selesi2024VirusDiffusion} and the present work delivers a comprehensive theoretical and computational analysis of its dynamics. Instead of analysing the fractional system directly, we exploit the analytical structure of the Caputo--Fabrizio operator to transform the fractional equations into an equivalent implicit system of ordinary differential equations. This transformation preserves the memory information contained in the original fractional model while replacing the fractional operator by additional derivative-correction terms. Consequently, the resulting system remains finite-dimensional but differs fundamentally from standard within-host compartmental models.
More precisely, the transformed model can be interpreted as a convex combination of two mechanisms governing the temporal evolution. The first component corresponds to the classical instantaneous within-host dynamics, whereas the second component represents a derivative-corrected contribution inherited from the Caputo--Fabrizio memory operator. The fractional order
$\mathfrak a\in[0,1]$ is therefore preserved as a model parameter controlling the relative contribution of these two mechanisms. This novel model given by equations \eqref{eq:H}--\eqref{eq:D} is the central object of our investigations. For $\mathfrak a=1$, the model reduces to the classical instantaneous system, while smaller values of
$\mathfrak a$ introduce progressively stronger derivative-correction effects originating from the underlying memory process.
This viewpoint differs substantially from the standard interpretation of fractional differential equations. Rather than studying the fractional operator itself, we investigate the qualitative properties of the equivalent derivative-corrected ordinary differential equation. The resulting implicit dynamical system belongs to a much less explored class of nonlinear differential equations, since the state derivatives appear on both sides of the model and become coupled through nonlinear interaction terms. As a consequence, many classical analytical techniques developed for explicit ordinary differential equations cannot be applied directly, creating new mathematical challenges concerning well-posedness, positivity, asymptotic behaviour and parameter sensitivity.

Despite the extensive literature devoted separately to within-host viral dynamics and fractional differential equations, comparatively little attention has been devoted to the mathematical analysis of implicit derivative-corrected systems obtained through transformations of fractional models. Most existing work concentrates either on numerical simulations or on the local stability of equilibria \cite{DingYe2009,%CardosoDosSantosCamargo2018,
	Moore2019%AdewoleFaniranAbdullahAli2023,AladeChumaJavedOlaniyiSangotolaGogovi2025
}, while rigorous investigations of existence, uniqueness, positivity, conservation laws, long-term dynamics and parameter dependence remain scarce. 

The purpose of the present paper is therefore to provide a comprehensive mathematical analysis of this derivative-corrected within-host viral dynamics model, and these are the main results obtained in Section~\ref{SEC:main}. We first rewrite the implicit system as an explicit vector field whenever the implicit matrix is invertible (Proposition~\ref{prop:explicit}) and establish sufficient conditions for local well-posedness and global continuation (Theorem~\ref{thm:lipsic}, Proposition~\ref{prop:suff}, and Remark~\ref{rem:global-existence}). We then prove positivity of all biological compartments and demonstrate conservation of the total cell population (Theorems~\ref{thm:positive-invariance} and~\ref{thm:conservation}), showing that the model remains biologically consistent despite its implicit structure. The asymptotic behaviour of the solutions is investigated in detail. We establish convergence of the infected-cell and viral populations to zero (Theorem~\ref{thm:decay}), derive the existence of the asymptotic healthy-cell population $H_\infty$, and obtain an explicit final-size relation describing the long-term tissue damage (Theorem~\ref{thm:final-size}). Furthermore, the basic reproduction number $R_0$ is computed using the next-generation matrix approach; the resulting expression is given in \eqref{eq:R0}. Interestingly, the threshold coincides with the classical reproduction number, indicating that the derivative correction modifies the transient dynamics rather than the invasion threshold itself. A particular emphasis is placed on the role of the derivative-correction parameter $\mathfrak a$. We investigate analytically how this parameter influences the temporal evolution of the infection, including the peak times and peak magnitudes of the infected-cell and viral populations, through the peak conditions \eqref{eq:Ipeak}--\eqref{eq:Vpeak}. Sufficient conditions ensuring monotone dependence on $\mathfrak a$ are derived and interpreted biologically in terms of the effective host response induced by the memory mechanism (Theorem~\ref{thm:early-monotonicity}). We further analyse the dependence of the final healthy-cell population on $\mathfrak a$ (Theorem~\ref{thm:final-size-monotonicity}), thereby quantifying the long-term protective effect of the derivative correction.

Finally, the theoretical analysis developed in Section~\ref{SEC:main} is complemented in Section~\ref{SEC:num} by numerical simulations illustrating the influence of the derivative-correction parameter on healthy cells, infected cells, viral load and cumulative cell death under biologically relevant parameter regimes. The numerical experiments illustrate the analytical results concerning peak infection times, peak-magnitude, viral-load dynamics, cumulative tissue damage and the dependence of the final healthy-cell population on the derivative-correction parameter $\ain$.

To the best of our knowledge, this is the first work providing a rigorous mathematical analysis of a derivative-corrected within-host viral dynamics model obtained from a Caputo--Fabrizio fractional formulation through transformation into an implicit system of ordinary differential equations. Beyond the particular application considered here, the analytical framework developed in this paper provides a general methodology for studying implicit nonlinear dynamical systems arising from fractional differential equations after elimination of the fractional operator.

\section{Basic notions}
For convenience of the readers who might stem from different fields we provide a short overview of the concepts (dynamical systems, matrices, implicit ODE, Caputo-Fabrizio calculus) that we use later on.

\subsection{Linearization and local stability}

In this paper we often investigate the local behaviour of nonlinear dynamical systems in a neighbourhood of their equilibrium points. For completeness, we briefly recall the standard linearization principle.

Consider the autonomous system
\begin{equation*}
	x'(t)=f(x(t)), \qquad x\in\mathbb{R}^n,
\end{equation*}
where
$f:\mathbb{R}^n\rightarrow\mathbb{R}^n$
is continuously differentiable. A point $x^\ast\in\mathbb{R}^n$ is called an equilibrium if
$f(x^\ast)=0$.

The Jacobian matrix of the vector field is defined by
\[
J(x)=Df(x)
=
\left(
\frac{\partial f_i}{\partial x_j}
\right)_{i,j=1}^n.
\]

Linearization of the nonlinear system at the equilibrium
$x^\ast$
leads to the linear system
\begin{equation*}
	y'(t)=J(x^\ast)y(t),
\end{equation*}
whose qualitative behaviour is completely determined by the spectrum
$\sigma(J(x^\ast)) = \{\lambda_1,\ldots,\lambda_n\}$. 	If every eigenvalue of
$J(x^\ast)$ has strictly negative real part, then
$x^\ast$ is locally asymptotically stable. If at least one eigenvalue has positive real part, then
$x^\ast$ is unstable.

For additional information on dynamical systems and stability theory, interested readers are referred to \cite{Perko2001%Khalil2002,%HirschSmaleDevaney2013
}.

\subsection{Metzler matrices and positivity}

Positivity plays a fundamental role in mathematical biology since all population variables represent non-negative quantities. A matrix
	$A=(a_{ij})\in\mathbb{R}^{n\times n}$
	is called a Metzler matrix if all its off-diagonal entries are non-negative, that is,
$a_{ij}\ge0$, $	i\neq j$. Metzler matrices characterize positive linear dynamical systems.
If the linear system
	\[
	x'(t)=Ax(t),
	\qquad
	x(0)\ge0.
	\]
is governed by a Metzler matrix	$A$, then $x(t)\ge0,$ for all 	$t\ge0.$

	Consequently, the positive orthant
	$\mathbb{R}_+^n$
	is positively invariant under the generated flow.

\subsection{The next-generation matrix}

One of the principal threshold quantities in mathematical epidemiology is the basic reproduction number $R_0$. In this paper it will be computed by means of the next-generation matrix method introduced by Diekmann \cite{Diekmann1990} and further developed by van den Driessche and Watmough \cite{VanDenDriesscheWatmough2002}. Suppose that the infection-related variables satisfy a system of the form
\begin{equation*}
	x'(t)=F(x)-V(x),
\end{equation*}
where $F$ represents the rates of appearance of new infections and $V$ collects all remaining transition terms. Linearizing the system at the disease-free equilibrium yields the matrices
\[
F
=
DF(x^\ast),
\qquad
V
=
DV(x^\ast).
\]

The next-generation matrix is then defined by $K=FV^{-1},$ provided that $V$ is invertible.
The basic reproduction number is defined as the spectral radius of the next-generation matrix,
	\[
	R_0
	=
	\rho(K)
	=
	\rho(FV^{-1}).
	\]

The quantity $R_0$ measures the expected number of newly infected cells generated by a single infected cell introduced into an otherwise infection-free environment. 

\subsection{Implicit ordinary differential equations}

Unlike classical compartmental models, the system studied in this paper naturally appears as an implicit system of ordinary differential equations. We therefore briefly recall the basic local solvability result.

Consider an implicit differential equation of the form
\begin{equation*}
	F(x,x')=0,
\end{equation*}
where $F:\mathbb{R}^n\times\mathbb{R}^n\longrightarrow\mathbb{R}^n$
is continuously differentiable.

The principal question is whether the equation can be rewritten locally as an explicit dynamical system, i.e. an ODE in normal form.

The following consequence of the Implicit Function Theorem provides the answer.	Suppose that
$F(x_0,y_0)=0$	and that the Jacobian matrix $D_{y}F(x_0,y_0)$	is invertible.
	Then there exist neighbourhoods	$U$	of $x_0$ and $V$ of	$y_0$,	together with a unique continuously differentiable mapping	$G:U\rightarrow V$, such that 
	\[
	F(x,G(x))=0,
	\qquad
	x\in U.
	\]
	Consequently, the implicit equation is locally equivalent to the explicit system
	\[
	x'(t)=G(x(t)).
	\]

This result constitutes the mathematical foundation of the existence and uniqueness analysis developed later.

For additional information on implicit ODEs and differential-algebraic equations, interested readers are referred to \cite{%HairerWanner1996,
KunkelMehrmann2006}.

\subsection{Caputo-Fabrizio fractional derivatives}

For a real absolutely continuous function $f$ the Caputo--Fabrizio \cite{CaputoFabrizio2015} fractional derivative of order $0<\mathfrak{a}<1$, is given by
\begin{equation}
	{}^{CF}D^{\mathfrak{a}}f(t)=\dfrac{1}{1-\mathfrak{a}}\int_{0}^{t}f'(\tau)e^{-\frac{\mathfrak{a}}{1-\mathfrak{a}}(t-\tau)}d\tau,
\end{equation}
while the corresponding fractional integral is
\begin{equation}\label{CFIntegral}
	{}^{CF}I^{\mathfrak{a}}f(t)=(1-\mathfrak{a})[f(t)-f(0)]+\mathfrak{a}\int_{0}^{t}f(\tau)d\tau.
\end{equation}
The relation between them is given by
\begin{align}\label{CFRelation1}
	{}^{CF}I^{\mathfrak{a}}\Big({}^{CF}D^{\mathfrak{a}}f(t) \Big)&=f(t)-f(0),\\ \label{CFRelation2}
	{}^{CF}D^{\mathfrak{a}}\Big({}^{CF}I^{\mathfrak{a}}f(t) \Big)&=f(t)-f(0)e^{-\frac{\mathfrak{a}}{1-\mathfrak{a}}t}.
\end{align}
The non-singular exponential kernel of the Caputo--Fabrizio fractional derivative constitutes one of its most attractive features for biological modelling. Unlike classical fractional derivatives with singular power-law kernels, the Caputo--Fabrizio operator represents a fading-memory mechanism in which the influence of past states decreases continuously with time. Consequently, recent events exert a stronger influence on the present dynamics than remote ones, a behaviour that is often more consistent with biological processes such as viral replication, cellular response and tissue adaptation. Mathematically, the Caputo--Fabrizio derivative can be viewed as a convolution of the ordinary derivative with an exponential kernel, thereby providing a natural description of finite-memory effects while avoiding the singularity associated with classical fractional operators. Although its mathematical properties have been the subject of ongoing discussion in the fractional calculus literature, the Caputo--Fabrizio operator has proved to be a useful modelling tool in numerous applications, where it successfully captures memory-dependent dynamics and offers an analytically tractable framework for incorporating hereditary effects into differential equation models.

Since $\lim_{t\rightarrow0+}{}^{CF}D^{\mathfrak{a}}f(t)=0$ for all admissible functions, fractional ODEs of the form ${}^{CF}D^{\mathfrak{a}}y(t)=g(t)$ need to satisfy $g(0)=0$ as a compatibility criterion for well-posedness. For $g(0)\neq0$ equations can be regularized by ${}^{CF}D^{\mathfrak{a}}y(t)=g(t)-g(0)e^{-\frac{\mathfrak{a}}{1-\mathfrak{a}}t}$, as it was done in \cite{SelesiTosic2025}.
Note that the regularization function is also in the kernel of the integral operator, i.e.,
\begin{equation}\label{CFkernel}
	{}^{CF}I^{\mathfrak{a}}\left(e^{-\frac{\mathfrak{a}}{1-\mathfrak{a}}t}\right)=0.	
\end{equation}
This term introduces an effect that vanishes as $t \to \infty$, so that the regularization has a larger effect at
the initial time $t = 0$, when the Caputo-Fabrizio derivative must be equal to zero, while at later time points this
effect becomes negligible and the function $g(t)$ takes over to have greater influence on the equation.

For additional information on Caputo--Fabrizio fractional calculus, especially its proper usage via regularization near $t=0$, interested readers are referred to \cite{AtanackovicPilipovicZorica2018,LosadaNieto2015,ZhangHuSun2018,SelesiTosic2025}.

\section{The model}\label{SEC:main}

Let $H, I, D$ denote the number of healthy or susceptible target cells\footnote{Although $T(t)$ is commonly used for uninfected target cells in the within-host literature, we retain the notation $H(t)$ to keep the compartment labels mnemonic and readily distinguishable in displayed formulas, oral presentation, and assistive reading, where $T(t)$ may be rendered ambiguously. Throughout the paper, the expression “healthy cells” is used solely as a concise synonym for “uninfected target cells” and does not imply any broader physiological characterization of these cells.}, infected cells and damaged, dead or removed tissue cells at a certain time. Let $V$ denote the number of virus particles at the same time. A simple fractional ODE model that describes their interaction, as a generalization of \cite{Baccam2006} and \cite{Rost2021} reads as:

\begin{align}\label{eq:odj}
	{}^{CF}D^\mathfrak{a} H(t)&= -b V(t)H(t) +b V(0)H(0)e^{-\frac{\mathfrak{a}}{1-\mathfrak{a}}t}\nonumber\\
	{}^{CF}D^\mathfrak{a} I(t)&= b V(t)H(t) - d I(t) - (b V(0)H(0)-d I(0))e^{-\frac{\mathfrak{a}}{1-\mathfrak{a}}t}\nonumber\\
	{}^{CF}D^\mathfrak{a} V(t)&= p I(t) - m V(t) - (p I(0)-m V(0))e^{-\frac{\mathfrak{a}}{1-\mathfrak{a}}t}\\
	{}^{CF}D^\mathfrak{a} D(t)&= d I(t) - d I(0)e^{-\frac{\mathfrak{a}}{1-\mathfrak{a}}t}\nonumber
\end{align}
with fractional derivative order $\mathfrak{a}\in(0,1)$ and parameters $b,d,p,m>0$ representing, respectively, the cell infection rate, cell death rate, cellular virus production rate, and viral clearance rate, together with the corresponding initial conditions
$$ H(0)=H_0\geq0, \quad I(0)=I_0\geq0, \quad V(0)=V_0\geq0, \quad D(0)=D_0\geq0.$$

The first step is to apply the fractional integral $I^\mathfrak{a}$ to the system \eqref{eq:odj}. We start with the first equation in \eqref{eq:odj}:

$$
I^{\alpha}(D^\mathfrak{a} H(t))=-b I^{\mathfrak{a}}  \left ( V(t)H(t) \right ) + bV(0)H(0)I_t^{\mathfrak{a}}\left ( e^{-\frac{\mathfrak{a}}{1-\mathfrak{a}}t} \right ) = -b I_t^{\mathfrak{a}}  \left ( V(t)H(t) \right ) .
$$
From \eqref{CFRelation2} we obtain
$$
H(t)-H(0)=-b\left [ (1-\mathfrak{a}) \left ( V(t)H(t)-V(0)H(0) \right ) +\mathfrak{a} \int_{0}^{t}V(s)H(s)ds\right ].
$$
Differentiating with respect to $t$ gives
$$
H'(t)=-b (1-\mathfrak{a})\left ( V(t)H(t) \right )' -b \mathfrak{a}V(t)H(t),
$$
i.e.,
$$
H'(t)=(1-\mathfrak{a}) \left [ -b V'(t)H(t)-b V(t)H'(t) \right ] +\mathfrak{a} \left [ -b V(t)H(t) \right ].
$$
Similarly, applying $I^\mathfrak{a}$ to the second equation in  \eqref{eq:odj} and using \eqref{CFRelation2}, we obtain:
\begin{eqnarray*}
I(t)-I(0) & = & (1-\mathfrak{a}) \left [ b \left ( V(t)H(t)-V(0)H(0) \right ) -d (I(t)-I(0) \right ] \\
 & + & b\mathfrak{a} \int_{0}^{t}V(s)H(s)ds - d\mathfrak{a} \int_{0}^{t}I(s)ds.
\end{eqnarray*}
Differentiating with respect to $t$ gives:
$$
I'(t) =  (1-\mathfrak{a}) \left [ b (V(t)H(t))'- d I'(t) \right ] +b \mathfrak{a} V(t)H(t) - \mathfrak{a}d I(t),
$$
i.e.,
\begin{eqnarray*}
I'(t) & = & (1-\mathfrak{a}) \left [ b V'(t)H(t)+b V(t)H'(t)- d I'(t) \right ] +\mathfrak{a} \left [ b V(t)H(t) - d I(t) \right ].
\end{eqnarray*}

For the virus equation, we first obtain:
\begin{eqnarray*}
V(t)-V(0) & = & (1-\mathfrak{a}) \left [ p I(t) - m V(t) -p I(0) + m V(0) \right ] \\
 & + & p\mathfrak{a} \int_{0}^{t}I(s) ds - m \mathfrak{a} \int_{0}^{t}V(s)ds.
\end{eqnarray*}
Differentiating with respect to $t$ gives:
%$$
%V'(t) =  (1-\mathfrak{a}) \left [ p I'(t) -m V'(t) \right ] +p \mathfrak{a} I(t) - m\mathfrak{a} V(t),
%$$
%i.e.,
\begin{eqnarray*}
V'(t) & = & (1-\mathfrak{a}) \left [  p I'(t) - m V'(t) \right ] +\mathfrak{a} \left [ p I(t) - m V(t) \right ].
\end{eqnarray*}

Finally, in a similar manner, we derive the ODE for $D(t)$. First, we obtain:
$$
D(t)-D(0)=(1-\mathfrak{a}) [d I(t) - d I(0)] + d \mathfrak{a} \int_{0}^{t}I(s) ds,
$$
and, after differentiating with respect to $t$:
$$
D'(t)=(1-\mathfrak{a}) dI'(t) +\mathfrak{a} d I(t).
$$

Thus, we arrive at a system of ODEs equivalent to \eqref{eq:odj}:

\begin{align}
	H'(t)&= \mathfrak{a} [-b V(t)H(t)]+(1-\mathfrak{a})[-b V'(t)H(t)-b V(t)H'(t)]\label{eq:H}\\
	I'(t)&= \mathfrak{a} [b V(t)H(t) - d I(t)]+(1-\mathfrak{a})[b V'(t)H(t)+b V(t)H'(t) - d I'(t)]\label{eq:I}\\
	V'(t)&= \mathfrak{a}[p I(t) - m V(t)] + (1-\mathfrak{a})[p I'(t) - m V'(t)]\label{eq:V}\\
	D'(t)&= \mathfrak{a} [d I(t)] + (1-\mathfrak{a})[d I'(t)]\label{eq:D}
\end{align}
with initial conditions
\begin{equation} H(0)=H_0, \quad I(0)=I_0, \quad V(0)=V_0, \quad D(0)=D_0.
\end{equation}

We note that the resulting model turns the right-hand side into a linear combination (more precisely, a convex combination with respect to $\mathfrak{a}$) of the original model, which contains zeroth-order derivatives on the right-hand side, and an enhanced model involving first-order derivatives of the classical model. We will refer to this system as the derivative-corrected within-host dynamical system. 

In the next sections we discuss the properties of this model from both mathematical and biological perspectives.

\begin{remark}[Dimensional consistency]
	Equations  \eqref{eq:H}--\eqref{eq:D} mix classical rate terms, such as $-bVH$, with derivatives of such rate terms, such as $(-bVH)'$. In dimensional variables these two terms have different physical units. Thus the model should either be interpreted in nondimensional time and nondimensional variables, or the coefficient multiplying the derivative-correction terms should contain a characteristic time scale. A dimensionally explicit variant would replace $1-\ain$ by a parameter with units of time, for example $\tau(1-\ain)$. 
	For additional information on correcting dimensional consistency in fractional and Caputo--Fabrizio-type models, and on parameter estimation from real viral-load datasets with physically meaningful time and concentration units, interested readers are referred to \cite{Baccam2006,CorreaEscuderoGomezAguilarLopezLopezAlvaradoMartinezBaleanu2022,HarroucheMomaniHasanAlSmadi2021,Yamaguchi2024}.

	In the present paper however we proceed in nondimensional variables.
\end{remark}

\subsection{Existence and uniqueness of a global solution}

Writing $X=(H,I,V,D)^T$, equations \eqref{eq:H}--\eqref{eq:D} can be written as
\[
M(H,V)X'=R(H,I,V),
\]
where
\[
M(H,V)=
\begin{pmatrix}
	1+(1-\ain)bV & 0 & (1-\ain)bH & 0\\
	-(1-\ain)bV & 1+(1-\ain)d & -(1-\ain)bH & 0\\
	0 & -(1-\ain)p & 1+(1-\ain)m & 0\\
	0 & -(1-\ain)d & 0 & 1
\end{pmatrix}
\]
and
\[
R(H,I,V)=
\begin{pmatrix}
	-\ain bVH\\
	\ain(bVH-dI)\\
	\ain(pI-mV)\\
	\ain dI
\end{pmatrix}.
\]
The determinant of the system matrix can be calculated as
\begin{equation}\label{eq:Delta}
	\Delta(H,V)=\det M(H,V)=(1+(1-\ain)d)(1+(1-\ain)m)(1+(1-\ain)bV)-(1-\ain)^2bpH.
\end{equation}

Thus, the next proposition shows that the system can be represented in its normal (explicit) form.

\begin{proposition}[Explicit vector field]\label{prop:explicit}
	If $\Delta(H,V)\ne0$, the implicit system is equivalent to the autonomous system $X'=f(X)$ where the explicit vector field is given by
	\begin{align}
		H'&=-\frac{\ain bH\bigl((1-\ain)pI+V(1+(1-\ain)d)\bigr)}{\Delta(H,V)},\label{eq:Hp}\\
		I'&=\frac{\ain\left[bH(V+(1-\ain)pI)-dI(1+(1-\ain)m)(1+(1-\ain)bV)\right]}{\Delta(H,V)},\label{eq:Ip}\\
		V'&=\frac{\ain\left[p\{I+(1-\ain)bV(H+I)\}-mV(1+(1-\ain)d)(1+(1-\ain)bV)\right]}{\Delta(H,V)},\label{eq:Vp}\\
		D'&=\frac{\ain d\left[(1-\ain)bHV+I(1+(1-\ain)m)(1+(1-\ain)bV)\right]}{\Delta(H,V)}.\label{eq:Dp}
	\end{align}
\end{proposition}

\begin{proof}
Since $M(H,V)$ is a nonsingular matrix, it follows that
\[
X'=M(H,V)^{-1}R(H,I,V),
\]
where
\[
R(H,I,V)
=
\ain
\begin{pmatrix}
	-bVH\\
	bVH-dI\\
	pI-mV\\
	dI
\end{pmatrix}.
\]
Computing the inverse matrix and using \eqref{eq:Delta} gives the system \eqref{eq:Hp}--\eqref{eq:Dp}.
\end{proof}

In the following we discuss the condition $\Delta(H,V)\ne0$ and state some sufficient conditions for it.

\begin{assumption}[Regularity region]\label{ass:regular}
	There exist a region $\Theta\subset\R_+^4$ and a constant $\eta>0$ such that
	\[
	\Delta(H,V)\ge \eta>0\qquad\text{for all }(H,I,V,D)\in\Theta.
	\]
\end{assumption}

\begin{theorem}[Local well-posedness]\label{thm:lipsic}
	Under Assumption \ref{ass:regular}, the vector field $f$ is locally Lipschitz on $\Theta$. Consequently, for every initial condition in $\Theta$, there exists a unique local solution as long as the trajectory remains in $\Theta$.
\end{theorem}

\begin{proof}
	The entries of $f$ from Proposition \ref{prop:explicit}	are rational functions whose only possible singularity in the nonnegative region is caused by $\Delta(H,V)=0$. If $\Delta\ge\eta>0$, then $f$ is continuously differentiable on compact subsets of $\Theta$, hence locally Lipschitz and Picard--Lindel\"of's theorem provides local existence and uniqueness.
\end{proof}

The following proposition provides a simple sufficient non-singularity condition that also describes the set of feasible parameters in the model. Since the cell infection rate $b$ and the virus production rate $p$ are typically small (actual values differ depending on measurement units \cite{ZitzmannKeRibeiroPerelson2024} but here we use the dimensionless model) compared with the initial healthy-cell population $H_0$, additionally damped by $(1-\ain)^2$  this condition is very mild. %If the condition is not met, then one would have to revert to a smaller number of initial $H_0$ and consider a locally smaller population of target cells.

\begin{proposition}[Sufficient regularity condition]\label{prop:suff}
	Assume $b,d,p,m\ge0$, $0\le \ain\le1$, and $0\le H(t)\le H_0$. If
	\begin{equation}\label{eq:suff_condition}
		(1-\ain)^2bp H_0 < (1+(1-\ain)d)(1+(1-\ain)m),
	\end{equation}
	then $\Delta(H(t),V(t))>0$ for every $V(t)\ge0$.
\end{proposition}

\begin{proof}
	Since $V\ge0$,
\begin{align*}
	\Delta(H,V)&=(1+(1-\ain)d)(1+(1-\ain)m)(1+(1-\ain)bV)-(1-\ain)^2bpH\\
	&\ge (1+(1-\ain)d)(1+(1-\ain)m)-(1-\ain)^2bpH.
\end{align*}
	Using $H\le H_0$ and \eqref{eq:suff_condition} gives $\Delta>0$.
\end{proof}

\begin{remark}[Global existence]\label{rem:global-existence}
	If the solution remains nonnegative and condition \eqref{eq:suff_condition} holds, then $H,I,D$ are bounded (this will follow from the conservation law that we will prove in the next section). Moreover, from \eqref{eq:Vp}, $V'$ is eventually negative and grows at most linearly for large $V$; therefore $V$ cannot blow up in finite time. Hence the solution extends globally. 
	%A completely minimal global-existence theorem would require a careful continuation argument near the surface $\Delta=0$, and is not attempted here.
\end{remark}

We also note that, since closed-form solutions are unavailable, solutions must be computed numerically. Care must be taken to avoid the region where $\Delta=0$, as numerical instabilities may otherwise occur.

\subsection{Positivity of the solutions and population conservation}
From a biological point of view, the number of healthy cells, infected cells, dead cells, as well as the number of virus particles need to be positive at all time points, hence we discuss these properties now. Also, the population size must be conserved at all time points. As we will see, the model satisfies all these properties.

\begin{theorem}[Positive invariance]\label{thm:positive-invariance}
	Assume $0<\ain\le1$, $b,d,p,m\ge0$, and $\Delta(H,V)>0$ along the solution. If the initial data are nonnegative, then
	\[
	H(t),I(t),V(t),D(t)\ge0
	\]
	for all $t\geq 0$ for which the solution exists.
\end{theorem}

\begin{proof}
	We check the vector field on the boundary of the nonnegative cone
\[
\mathcal C=
\left\{
(H,I,V,D)\in\mathbb{R}^4:
H,I,V,D \geq 0
\right\}.
\] 

From \eqref{eq:Hp}, $H'\leq0$, thus $H\leq H_0$, and if $H=0$, then $H'=0$. So $H$ cannot cross from positive to negative. From \eqref{eq:Ip}, if $I=0$, then
	\[
	I'=\frac{\ain bHV}{\Delta(H,V)}\ge0.
	\]
	From \eqref{eq:Vp}, if $V=0$, then
	\[
	V'=\frac{\ain pI}{\Delta(H,0)}\ge0.
	\]
	From \eqref{eq:Dp}, if $D=0$, then $D'$ is still independent of $D$ and satisfies
	\[
	D'=\frac{\ain d\left[(1-\ain)bHV+I(1+(1-\ain)m)(1+(1-\ain)bV)\right]}{\Delta(H,V)}\ge0.
	\]
	Thus the vector field is quasi-positive, and the nonnegative cone $\mathcal C$ is positively invariant.
\end{proof}

\begin{theorem}[Conservation law]\label{thm:conservation}
	The total number of cells is preserved, i.e.
	\[
	H(t)+I(t)+D(t)=const.
	\]
	for all $t\geq 0$ for which the solution exists.
\end{theorem}

\begin{proof}
Adding \eqref{eq:H}, \eqref{eq:I}, and \eqref{eq:D} gives $H'(t)+I'(t)+D'(t)=0$, which implies the exact conservation law
\begin{equation}\label{eq:conservation}
	H(t)+I(t)+D(t)=H_0+I_0+D_0=:N.
\end{equation}
Consequently $0\le H(t),I(t),D(t)\le N$ whenever the solution is nonnegative.
\end{proof}

\subsection{Long-time behavior}
We are also interested in the long-time behavior of the system: whether the numbers of different cell types and virus particles stabilize over time, and what the final sizes of the healthy, infected, and dead-cell populations are.

Throughout this section assume $0<\ain\le1$, $d,m>0$, nonnegative global solutions, and $\Delta(H,V)\ge\eta>0$ along the trajectory.

\begin{lemma}[Integrability of infected cells]\label{lem:Iint}
It holds that $I\in L^1(0,\infty)$.
\end{lemma}

\begin{proof}
	From \eqref{eq:D},
	\[
	D'=\ain dI+(1-\ain)dI'.
	\]
	Integrating from $0$ to $T$ yields
	\[
	D(T)-D_0=\ain d\int_0^T I(s)\,\dd s+(1-\ain)d(I(T)-I_0).
	\]
	Hence
	\[
	\ain d\int_0^T I(s)\,\dd s=D(T)-D_0-(1-\ain)d(I(T)-I_0).
	\]
	The right-hand side is bounded uniformly in $T$ because $D(T)$ and $I(T)$ are bounded by \eqref{eq:conservation}. Since $\ain d>0$, it follows that $I\in L^1(0,\infty)$.
\end{proof}

\begin{lemma}[Integrability of viral load]\label{lem:Vint}
It holds that $V\in L^1(0,\infty)$.
\end{lemma}

\begin{proof}
	From \eqref{eq:V},
	\[
	(1+(1-\ain)m)V'=\ain pI-\ain mV+(1-\ain)pI'.
	\]
	Integrating gives
	\[
	(1+(1-\ain)m)(V(T)-V_0)=\ain p\int_0^T I(s)\,\dd s-\ain m\int_0^T V(s)\,\dd s+(1-\ain)p(I(T)-I_0).
	\]
	Therefore
	\[
	\ain m\int_0^T V(s)\,\dd s=\ain p\int_0^T I(s)\,\dd s+(1-\ain)p(I(T)-I_0)-(1+(1-\ain)m)(V(T)-V_0).
	\]
	The right-hand side is bounded above uniformly in $T$: the first term is bounded by Lemma \ref{lem:Iint}, the second by boundedness of $I$, and the term $-(1+(1-\ain)m)V(T)$ is nonpositive. Since $\ain m>0$, it holds that $V\in L^1(0,\infty)$.
\end{proof}

\begin{lemma}[Boundedness and uniform continuity]\label{lem:bounded-uniform-continuity}
 $I$ and $V$ are uniformly continuous on $[0,\infty)$.
\end{lemma}

\begin{proof}
	The conservation law bounds $H,I,D$. The formula \eqref{eq:Vp} shows that $V'$ is negative for sufficiently large $V$ and has at most linear growth; hence $V$ is bounded. Since $\Delta\ge\eta>0$, the explicit vector field \eqref{eq:Hp}--\eqref{eq:Dp} is bounded on the bounded invariant region reached by the solution. Thus $I'$ and $V'$ are bounded, which implies uniform continuity.
\end{proof}

\begin{theorem}[Decay of infected cells and virus]\label{thm:decay}
	\[
	\lim_{t\to\infty}I(t)=0,
	\qquad
	\lim_{t\to\infty}V(t)=0.
	\]
\end{theorem}

\begin{proof}
	The functions $I$ and $V$ are nonnegative, integrable on $[0,\infty)$, and uniformly continuous. By Barbalat's lemma, both tend to zero as $t\to\infty$.
\end{proof}

The latter results show an important feature of the model: the viral load and the number of infected cells both decay to zero. As the analysis continues, we will see that this need not hold for the numbers of healthy and dead cells.

From \eqref{eq:Hp}, $H'(t)\le0$. Since $H(t)\ge0$, the limit
\[
H_\infty:=\lim_{t\to\infty}H(t)
\]
exists and is nonnegative. We now prove that it is strictly positive. So the susceptible/healthy cell population does not hit zero in finite time and does not even decay asymptotically to zero.

From the first equation,
\[
(1+(1-\ain)bV)H'+(1-\ain)bHV'=-\ain bVH.
\]
Therefore
\[
\frac{\dd}{\dd t}\log\bigl(H(1+(1-\ain)bV)\bigr)
=-\ain b\frac{V}{1+(1-\ain)bV}.
\]
Integrating from $0$ to $T$ gives
\[
H(T)(1+(1-\ain)bV(T))=H_0(1+(1-\ain)bV_0)
\exp\left(-\ain b\int_0^T\frac{V(s)}{1+(1-\ain)bV(s)}\,\dd s\right).
\]
Since $0\le V/(1+(1-\ain)bV)\le V$ and $V\in L^1(0,\infty)$, the integral has a finite limit. Since $V(T)\to0$, we obtain
\begin{equation}\label{eq:Hinf}
	H_\infty=H_0(1+(1-\ain)bV_0)
	\exp\left(-\ain b\int_0^\infty\frac{V(s)}{1+(1-\ain)bV(s)}\,\dd s\right)>0.
\end{equation}
Thus
\[
H(t)\downarrow H_\infty>0.
\]
By \eqref{eq:conservation} and $I(t)\to0$,
\[
D(t)\uparrow D_\infty=N-H_\infty.
\]

\begin{theorem}[Final-size identity]\label{thm:final-size} It holds that 
\begin{equation}\label{eq:final_general}
		\log\frac{H_0(1+(1-\ain)bV_0)}{H_\infty}
		=\ain b\int_0^\infty\frac{V(t)}{1+(1-\ain)bV(t)}\,\dd t.
\end{equation}	
\end{theorem}

\begin{proof}
Equation \eqref{eq:Hinf} yields the exact final-size identity \eqref{eq:final_general}.
\end{proof}

We note that this is closed only if the integral can be expressed in terms of $H_\infty$ and initial data.

For $\ain=1$,  the classical algebraic final-size equation is recovered. Indeed,
\[
\int_0^\infty V(t)\,\dd t=\frac{1}{m}\left(V_0+\frac{p}{d}(H_0+I_0-H_\infty)\right),
\]
so
\begin{equation}\label{eq:final_classical}
		\log\frac{H_0}{H_\infty}
		=\frac{b}{m}\left(V_0+\frac{p}{d}(H_0+I_0-H_\infty)\right).
\end{equation}
For a very small but nonzero initial infectious dose, neglecting $I_0$ and $V_0$ in \eqref{eq:final_classical} and writing $h_\infty=\frac{H_\infty}{H_0}$ gives
$$-\log h_\infty=R_0(1-h_\infty).$$
Thus, in the classical case, the surviving fraction of healthy cells is determined solely by $R_0$.

For $0<\ain<1$, one still has integral identities. From the dead-cell equation \eqref{eq:D},
\begin{equation}\label{eq:intI}
	\ain d\int_0^\infty I(t)\,\dd t=H_0+I_0-H_\infty+(1-\ain)d I_0.
\end{equation}
From the virus-equation \eqref{eq:V},
\begin{equation}\label{eq:intV}
	\ain m\int_0^\infty V(t)\,\dd t
	=\ain p\int_0^\infty I(t)\,\dd t-(1-\ain)p I_0+(1+(1-\ain)m)V_0.
\end{equation}
However, \eqref{eq:final_general} involves $V/(1+(1-\ain)bV)$, not $V$ itself. Thus the modified model generally does not yield the same closed algebraic final-size equation. The dependence of $H_\infty$ on $\ain$ is therefore genuinely $\ain$-dependent. The integral identity shows that $\ain$ affects final size not only through timing, but also through an effective saturation of viral exposure.

\subsection{Extrema of the solution}
We now analyze the infection peak times and peak sizes, as well as the viral-load peak time and peak size.

Since $H'(t)\le0$,
\[
\max_{t\ge0}H(t)=H_0,
\qquad
\inf_{t\ge0}H(t)=H_\infty.
\]
Because $D'(t)\ge0$,
\[
\min_{t\ge0}D(t)=D_0,
\qquad
\sup_{t\ge0}D(t)=D_\infty.
\]
Interior extrema of $I$ satisfy $I'=0$, equivalently $I(t)$ has a local maximum at any time $t_I^*$
satisfying
\begin{equation}\label{eq:Ipeak}
	bH(t_I^*)(V(t_I^*)+(1-\ain)pI(t_I^*))=dI(t_I^*)(1+(1-\ain)m)(1+(1-\ain)bV(t_I^*)).
\end{equation}
Interior extrema of $V$ satisfy $V'=0$, equivalently $V(t)$ has a local maximum at any time $t_V^*$
\begin{equation}\label{eq:Vpeak}
	p\{I(t_V^*)+(1-\ain)bV(t_V^*)(H(t_V^*)+I(t_V^*))\}=mV(t_V^*)(1+(1-\ain)d)(1+(1-\ain)bV(t_V^*)).
\end{equation}
In general, $t_I^*$ and  $t_V^*$ as well as $I_{\max}=I(t_I^*)$ and $V_{\max}=V(t_V^*)$
cannot generally be calculated explicitly in closed form and must instead be estimated numerically by solving the ODE system.

%Only with additional assumptions: regularity of the solutions, that  $I$ and $V$ are unimodal, the peaks are nondegenerate, and the sensitivity functions satisfy the appropriate sign conditions (needed for the implicit theorem) one can prove that peak times and peak heights are monotone with respect to $\ain$. But there is no reason that the sign conditions should automatically hold.

For $\ain=1$, \eqref{eq:Ipeak} and \eqref{eq:Vpeak}  reduce to the classical conditions
\[
bH(t_I^*)V(t_I^*)=dI(t_I^*),
\qquad
pI(t_V^*)=mV(t_V^*).
\]
For $\ain<1$, the peak hypersurfaces are shifted by the derivative-correction term $(1-\ain)$. Indeed, Proposition \ref{prop:explicit} shows that the system has the matrix form $$M(H,V,1-\ain)X'=\ain R(H,I,V),$$ where $R$ does not depend on $\ain$. Hence, if $\ain=0$ and $M$ is nonsingular, the system reduces to $X'=0$ and the solution is frozen at the initial state $X(t)=X_0$. Therefore, as $\ain\to 0$, the infection dynamics become slower; in particular, smaller $\ain$ delays the rise of $I$ and $V$. However, this is not merely a time rescaling, because $M(H,V,1-\ain)$ also depends on $1-\ain$. Thus, $\ain$ changes both the speed of evolution and the shape of the vector field. Equations \eqref{eq:Ipeak} and \eqref{eq:Vpeak} show that $\ain$ affects the peak locations through the shifting factor $1-\ain$. Consequently, $\ain$ can affect both peak time and peak height.

\subsection{Basic reproduction number}

We can compute $R_0$ from the disease-free equilibrium by linearizing the infection-driving variables $(I,V)$ around $I=V=0$, with $H=H_0$ since $D$ is only an accumulated removed/dead compartment.

Near $(H_0,0,0,D_0)$, the infection-related variables are $x=(I,V)^T$. The new-infection and transition matrices in the classical infected subsystem are
\[
F=\begin{pmatrix}0&bH_0\\0&0\end{pmatrix},\qquad
\mathcal V=\begin{pmatrix}d&0\\-p&m\end{pmatrix}.
\]
Thus
\[
\mathcal V^{-1}=\begin{pmatrix}1/d&0\\p/(dm)&1/m\end{pmatrix},
\qquad
K=F\mathcal V^{-1}=\begin{pmatrix}bp H_0/(dm)&b H_0/m\\0&0\end{pmatrix}.
\]
Therefore
\begin{equation}\label{eq:R0}
	R_0=\rho(K)=\frac{bp H_0}{dm}.
\end{equation}

\begin{remark}[Why $\ain$ does not appear in $R_0$]
	The implicit linearization has the form
	\[
	A x'=\ain(F-\mathcal V)x,
	\]
	where
	\[
	A=\begin{pmatrix}1+(1-\ain)d&-(1-\ain)bH_0\\-(1-\ain)p&1+(1-\ain)m\end{pmatrix}.
	\]
	The matrix $A$ changes transient rates and eigenvalues away from the zero crossing, but the threshold at which a zero eigenvalue appears is determined by
	\[
	\det(F-\mathcal V)=0,
	\]
	provided $A$ is nonsingular. This gives $bp H_0=dm$, equivalently $R_0=1$. Thus $\ain$ does not alter the local invasion threshold, although it can alter peak time, peak height, duration, and final size.
\end{remark}

The local disease-free threshold at $(H_0,0,0,D_0)$ is governed by $R_0$. If $R_0<1$, the linearized infected subsystem is locally stable, and sufficiently small infections decay. If $R_0>1$, the disease-free equilibrium with target-cell level $H_0$ is unstable to viral invasion.

More generally, every disease-free equilibrium $(H_*,0,0,D_*)$ has local threshold
\[
R_0(H_*)=\frac{bpH_*}{dm}.
\]
During the infection, $H(t)$ decreases. The critical target-cell level
\[H_c=\frac{dm}{bp}\]
marks the stability threshold for the family of disease-free equilibria. Accordingly, the quantity $R_{\mathrm{eff}}(t)=bpH(t)/(dm)$ may be interpreted as a frozen-state reproduction indicator: once $H(t)<H_c$, it falls below one, meaning that a sufficiently small infection would decay if the current target-cell level were held fixed. This crossing need not coincide with the peaks of $I(t)$ or $V(t)$, since infected cells and viral particles already present may sustain further transient growth; their eventual decay follows from Theorem~\ref{thm:decay}. The parameter $\ain$ changes the time scale and transient shape of the infection, but not the equilibrium threshold $H_c$ separating local invasion from local decay near a disease-free state.

\begin{remark}
	The independence of $R_0$ from $\ain$ is not contradictory to the dependence of $H_\infty$ on $\ain$. The number $R_0$ is a local invasion threshold at the disease-free equilibrium, whereas $H_\infty$ is a global outcome depending on the entire nonlinear trajectory.
\end{remark}

For a complete stability and bifurcation analysis we compute the linearized system and its eigenvalues.
Consider a disease-free equilibrium
$$
E_{H_*}=(H_*,0,0,D_*).
$$
The equilibria form a continuum:
$$\mathcal E_0={(H_*,0,0,D_*):H_*,D_*\ge0}.$$

The relevant threshold at $E_{H_*}$ is $R_0(H_*)=\frac{bpH_*}{dm}$.

Near $E_{H_*}$, the infection--describing variables are $(I,V)$. Their linearized system is
\begin{equation}\label{eq:ABs}
A(\ain)
\begin{pmatrix}
	I'\\
	V'
\end{pmatrix}
=
\ain B
\begin{pmatrix}
	I\\
	V
\end{pmatrix},
\end{equation}
where
\begin{equation}\label{eq:AB}
A(\ain)=
\begin{pmatrix}
	1+(1-\ain)d & -(1-\ain)bH_*\\
	-(1-\ain)p & 1+(1-\ain)m
\end{pmatrix},
\qquad \mbox{ and }\qquad
B=
\begin{pmatrix}
	-d & bH_*\\
	p & -m
\end{pmatrix}.
\end{equation}

The determinant of matrix $A$ is

$$\Delta_*= (1+(1-\ain)d)(1+(1-\ain)m)-(1-\ain)^2bpH_*.$$

From Assumption \ref{ass:regular} we have 
$\Delta_*>0$ so that the linearized implicit system is regular.

The characteristic equation is $det(\lambda A-\ain B)=0.$ 
A useful simplification is obtained by setting 
$\mu=\frac{\lambda}{\ain+(1-\ain)\lambda}.$

Then $\mu$ satisfies the classical characteristic equation: 
\begin{equation}\mu^2+(d+m)\mu+dm(1-R_0(H_*))=0.\label{eq:mu}
\end{equation}

Thus the implicit eigenvalues $\lambda$ of our model are related to the classical eigenvalues $\mu$ by
\begin{equation}\lambda=\frac{\ain\mu}{1-(1-\ain)\mu}.\label{eq:eigenvalue}
\end{equation}

The classical infection eigenvalues $\mu_1,\mu_2$ corresponding to $I,V$ satisfy
\eqref{eq:mu}. Therefore, one can distinguish the following cases:

Case 1: $R_0(H_*)<1$. Both $\mu_1,\mu_2<0$. Since $1-(1-\ain)\mu_j>0$, we get
$\lambda_j<0$, $j=1,2$. So $E_{H_*}$ is locally stable in the infected directions, meaning 
infection dies out locally.

Case 2: $R_0(H_*)>1$. One classical eigenvalue $\mu_1$ is positive and the other one $\mu_2$ is negative. Under the regularity condition $\Delta_*>0$, the positive eigenvalue remains positive after the transformation $\lambda=\frac{\ain\mu}{1-(1-\ain)\mu}.$ Indeed, $\Delta_*=\bigl[1-(1-\ain)\mu_1\bigr]\bigl[1-(1-\ain)\mu_2\bigr]$. Since $\mu_2<0$ the second factor is positive; hence $\Delta_*>0$ implies $1-(1-\ain)\mu_1>0$, and therefore $\lambda_1=\ain\mu_1/[1-(1-\ain)\mu_1]>0$.
Thus $E_{H_*}$ is unstable in the infected directions, meaning the infection can invade.

Case 3: $R_0(H_*)=1$. One eigenvalue is $\mu_1$ zero and the other one $\mu_2$ is negative, so that $\lambda_1=0$, too. Therefore the equilibrium is nonhyperbolic.
This is the threshold case.

The critical target-cell level is $H_c=\frac{dm}{bp}$. Equilibria with $H_*<H_c$ are stable in infected directions, while equilibria with $H_*>H_c$ are unstable in infected directions.

The parameter $\ain$ does not change the stability threshold, i.e. 
$R_0(H_*)=1$ is independent of $\ain$, but it changes the eigenvalues $\lambda_j$ via \eqref{eq:eigenvalue}. So $\ain$ changes the speed of growth or decay. As $\ain\to 0$, we get $\lambda_j\to 0$, $j=1,2$. Thus the dynamics become slower and approach the frozen case.
This means that $\ain$ does not decide whether infection invades, but controls how fast it unfolds.

Near the disease-free state, the dominant infected eigenvalue is $\lambda_+(\ain) = \frac{\ain\mu_+}{1-(1-\ain)\mu_+}$, where $\mu_+$ is the positive classical eigenvalue when $R_0>1$.
If $0<\mu_+<1$, then $\frac{d\lambda_+}{d\ain} >0$. Thus increasing $\ain$ increases the initial exponential growth rate governed by $\lambda_+$; while decreasing $\ain$ slows early infection.
Therefore, under early-stage approximation and unimodality, smaller $\ain$ means later infection peaks.

Concerning the bifurcation structure we see there is a threshold at
$R_0(H_*)=1$. However, this is not a standard endemic transcritical bifurcation in the closed target-cell model, because there is no positive endemic equilibrium with $I_*>0,\, V_*>0.$ Indeed, at equilibrium, $D'=dI$ in the classical part, and the implicit terms vanish when derivatives vanish. Thus, for $d>0$, $I_*=0.$ Then $V_*=0$.
Therefore the only equilibria are disease-free equilibria.

The infected characteristic equation for $\mu$ is quadratic:
$\mu^2+(d+m)\mu+dm(1-R_0)=0.$ Its discriminant is
$(d+m)^2-4dm(1-R_0) = (d-m)^2+4dmR_0>0.$ Therefore the classical eigenvalues are real. Due to \eqref{eq:eigenvalue} the implicit eigenvalues $\lambda$ are also real as long as the system is regular. Thus, no Hopf bifurcation occurs at the disease-free equilibrium.

There is another mathematical threshold: $$\Delta_*=(1+(1-\ain)d)(1+(1-\ain)m)-(1-\ain)^2bpH_*=0.$$
At this point the implicit system matrix $A$ becomes singular. This is not a biological infection threshold; it is a singularity of the implicit formulation. Imposing Assumption \ref{ass:regular} we have avoided this singularity and made the model biologically meaningful.

\subsection{Dependence on the order of fractional derivative $\ain$}
\subsubsection{Global effects of $\ain$}
The parameter $\ain$ affects the model in two distinct ways. First, every explicit derivative contains the factor $\ain$, so as $\ain\downarrow0$ the dynamics slow down. Secondly, the vector field also depends on $1-\ain$, so changing $\ain$ is not a pure rescaling of time. This means that $\ain$ can change peak timing, peak magnitude, duration, and final size.

On every compact region on which $\Delta\ge\eta>0$, the vector field $f$ from Proposition \ref{prop:explicit}  satisfies
\[
\|f_\ain(X)\|\le C\ain
\]
for $\ain$ small, with $C$ depending on the compact set and the remaining parameters $b,p,d,m>0$. Thus, on fixed finite time intervals, solutions remain closer to their initial state as $\ain\to0$. This supports the interpretation that smaller $\ain$ delays infection dynamics. 

%A global monotonic theorem for peak time as a function of $\ain$ would require additional assumptions such as unimodality of $I$ and $V$ and uniform regularity of the peak hypersurfaces \eqref{eq:Ipeak}--\eqref{eq:Vpeak}.

The parameter $\ain$ influences also the infection outcome. The final size $H_\infty$ depends on $\int_0^\infty\frac{V(t)}{1+(1-\ain)bV(t)}\,\dd t$ i.e. on how high the viral peak becomes,
how long the virus persists, and the entire nonlinear course of infection.

\subsubsection{Sensitivity and monotonicity with respect to $\ain$}\label{subsec:monoton}
Next, we discuss monotonicity of the infected cells $I(t,\ain)$, viral load $V(t,\ain)$ and final size $H_\infty(\ain)$ under additional sufficient assumptions. These are both mathematically rigorous and biologically feasible for acute, single-wave infections where most target-cell loss occurs near the growth and peak phase. Single-wave behaviour motivates part of the biological discussion, but uniqueness of the viral peak is not assumed in the final-size monotonicity theorem.

Recall that under Assumption \ref{ass:regular} the implicit system admits the equivalent explicit formulation \[X'=F(X,\mathfrak a),
\qquad
X=(H,I,V,D),
\] and the solution remains in the positive invariant region $X\geq 0$.
Since the vector field is continuously differentiable with respect to both
the state variables and the parameter $\mathfrak a$, the solution depends
continuously and differentiably on $\mathfrak a$.

Define the sensitivity vector
\[
Z(t)
=
\frac{\partial X(t,\mathfrak a)}
{\partial\mathfrak a}
=
(H_{\mathfrak a},
I_{\mathfrak a},
V_{\mathfrak a},
D_{\mathfrak a})^T.
\]

Differentiating the explicit system with respect to $\mathfrak a$ yields the
variational system
\[Z'=
D_XF(X,\mathfrak a)Z
+\partial_{\mathfrak a}F(X,\mathfrak a),
\]
with homogeneous initial condition $Z(0)=0.$

Consequently,
\[
Z(t)=\int_0^t
U(t,s)\,\partial_{\mathfrak a}F(X(s),\mathfrak a)\,ds,
\]
where $U(t,s)$ denotes the fundamental matrix associated with the linearized
system. Therefore the sensitivity of each state variable is completely determined by
the variational equation. A global result determining the sign of $\partial_{\mathfrak a}F$ for all $t\geq 0$ seems out of reach; therefore, we look for local results near the initial conditions.

Assume that the infection is in its early stage and target-cell depletion is
still negligible,
\[H(t)\approx H_0.\]
Then the infected subsystem $(I,V)$ is approximated by \eqref{eq:ABs} where the system matrices are \eqref{eq:AB} with $H_*$ replaced by $H_0$.

Since $\Delta_\ain(H_0,0)>0$, 
\begin{equation*}
	\begin{pmatrix}
		I'\\
		V'
	\end{pmatrix}
	=
	\ain 	A(\ain)^{-1} B
	\begin{pmatrix}
		I\\
		V
	\end{pmatrix} = C(\ain)	\begin{pmatrix}
	I\\
	V
\end{pmatrix},
\end{equation*}
with $C(\ain) = \ain 	A(\ain)^{-1} B$. A direct calculation provides
\begin{equation*}
C(\ain) = \frac{\ain}{\Delta_\ain}	
\begin{pmatrix}
	(1-\ain)bpH_0-d(1+(1-\ain)m) & bH_0\\
p & (1-\ain)bpH_0-m(1+(1-\ain)d)
	\end{pmatrix}.
\end{equation*}
The off-diagonal entries are positive:
$C_{12}=\frac{\ain bH_0}{\Delta_\ain}>0$ and
$C_{21}=\frac{\ain p}{\Delta_\ain}>0$; hence $C(\ain)$ is a Metzler matrix and the linear system is cooperative.

\begin{theorem}[Conditional early-stage comparison]\label{thm:early-monotonicity}
	Suppose $0<\mathfrak a_1<\mathfrak a_2\le1$, and assume that
$C(\mathfrak a_2)-C(\mathfrak a_1)$	is entrywise nonnegative.
	
Then, for identical positive initial data,
	\[
	I(t,\mathfrak a_1)
	\le
	I(t,\mathfrak a_2), \qquad \mbox{ and }\qquad 
		V(t,\mathfrak a_1)
	\le
	V(t,\mathfrak a_2),
	\]
for every time during which the early-stage approximation remains valid.
	
\end{theorem}

\begin{proof}
	
	Since the coefficient matrices $C(\ain_1)$ and $C(\ain_2)$  are Metzler matrices, and $C(\ain_2)\geq C(\ain_1)$ entrywise, the comparison theorem for cooperative linear systems gives
\[	\begin{pmatrix}
	I\\
	V
\end{pmatrix}(t,\mathfrak a_1)
	\le
		\begin{pmatrix}
		I\\
		V
	\end{pmatrix}(t,\mathfrak a_2),
\]
which immediately yields the desired inequalities.
	
\end{proof}

The comparison theorem shows that increasing the parameter
$\mathfrak a$ increases the early amplification of both infected cells and
viral particles. Equivalently, decreasing $\ain$ delays the rise of the infection.

This agrees with the interpretation of $\mathfrak a$ as a parameter
controlling the balance between instantaneous dynamics and derivative-induced
memory effects.

The condition $C(\mathfrak a_2)-C(\mathfrak a_1)\geq 0$ is sufficient, but not necessary, for the stated componentwise ordering. It is stronger than cooperativity of each $C(\ain)$. For the baseline parameters used in Section~\ref{SEC:num}, the off-diagonal entries increase with $\ain$, whereas the diagonal entries become more negative, so this entrywise matrix-ordering condition does not hold. Consequently, Theorem~\ref{thm:early-monotonicity} is not invoked to justify the baseline simulations; failure of its sufficient condition does not imply failure of the monotonicity patterns observed numerically.

\begin{remark}
These results show that we have $\frac{d}{d\ain}V(t,\ain)\geq 0$ before the viral peak $t\leq t_V^*$ under weak depletion and cooperativity. After the peak, however, one should not expect $\frac{d}{d\ain}V(t,\ain)\geq 0$ globally. A larger $\ain$ may produce a higher and earlier peak and then a faster decline, so $\frac{d}{d\ain}V(t,\ain)$ may change sign later.
\end{remark}

Concerning the final healthy-cell population size, we recall from \eqref{eq:Hinf} that the
exact final-size relation is
\[
H_\infty(\mathfrak a)
=
H_0
\left(
1+(1-\mathfrak a)bV_0
\right)
\exp
\left(
-
\mathfrak a b
\int_0^\infty
\frac{V(t,\mathfrak a)}
{1+(1-\mathfrak a)bV(t,\mathfrak a)}
\,dt
\right).
\]

Define the effective cumulative viral exposure

\[
\mathcal E(\mathfrak a)
=
\mathfrak a
\int_0^\infty
\frac{V(t,\mathfrak a)}
{1+(1-\mathfrak a)bV(t,\mathfrak a)}
\,dt.
\]

\begin{theorem}[Conditional final-size monotonicity]\label{thm:final-size-monotonicity}
	Let $0<\mathfrak a_1<\mathfrak a_2\leq1$, with the same model parameters and identical nonnegative initial data, and assume the standing hypotheses ensuring the final-size identity for both solutions. If
	\[
	\mathcal E(\mathfrak a_1)
	\leq
	\mathcal E(\mathfrak a_2),
	\]
	then
	\[
	H_\infty(\mathfrak a_1)
	\geq
	H_\infty(\mathfrak a_2).
	\]
\end{theorem}

\begin{proof}
	Since $bV_0\geq0$ and $\mathfrak a_1<\mathfrak a_2$, the prefactor in the final-size identity satisfies
	\[
	1+(1-\mathfrak a_1)bV_0
	\geq
	1+(1-\mathfrak a_2)bV_0.
	\]
	Moreover, the assumed ordering of the effective exposures gives
	$\exp[-b\mathcal E(\mathfrak a_1)]\geq
	\exp[-b\mathcal E(\mathfrak a_2)]$. Equivalently, the exact final-size identity yields
	\[
	\frac{H_\infty(\mathfrak a_1)}{H_\infty(\mathfrak a_2)}
	=
	\frac{1+(1-\mathfrak a_1)bV_0}
	{1+(1-\mathfrak a_2)bV_0}
	\exp\!\left(b\left[\mathcal E(\mathfrak a_2)-\mathcal E(\mathfrak a_1)\right]\right)
	\geq1,
	\]
	which proves the claim.
\end{proof}

\begin{remark}
	The previous theorem states that if increasing
	$\mathfrak a$ increases the effective cumulative viral burden, then the final
	healthy-cell population decreases monotonically.
	Biologically, this means that stronger instantaneous infection dynamics lead
	to greater cumulative tissue damage, whereas smaller values of
	$\mathfrak a$ preserve a larger fraction of healthy target cells by slowing
	the overall infection dynamics.
	
	The result that $\frac{d}{d\ain}H_\infty(\ain)<0$ is also conditioned on the assumption that effective cumulative viral exposure $\mathcal E(\mathfrak a)$ is increasing, while total viral exposure may fail to be monotone and hence an unconditional result is unlikely. Even if larger $\ain$ causes a faster and higher early viral peak, it may also produce a faster target-cell depletion, and hence a faster subsequent collapse of viral production. Therefore total viral exposure could theoretically fail to be monotone in $\ain$.
\end{remark}

\begin{remark}
We elaborate further on the condition that the effective viral exposure $\mathcal E(\mathfrak a)$ is monotone increasing in $\ain$. If we denote the integrand by \[G(\ain, V)=\frac{\ain V}
{1+(1-\ain)bV},\] then $\mathcal E(\mathfrak a)=\int_0^\infty G(\ain, V(t,\ain))dt$. Since $\frac{d}{d\ain}G\geq0$ and $\frac{d}{d V}G\geq0$, $G$ is increasing in both $\ain$ and $V$.
Therefore, a condition of the form $$\ain_1<\ain_2 \qquad\mbox{ and }\qquad V(t,\ain_1)\leq V(t,\ain_2),\quad t\geq 0$$ would imply $$G(\ain_1, V(t,\ain_1))\leq G(\ain_2, V(t,\ain_2)), \;t\geq 0$$ pointwise and hence integration would yield $\mathcal E(\ain_1)\leq \mathcal E(\ain_2)$. However, this global pointwise ordering of $V$ is biologically unlikely after the peak, because larger $\ain$ may produce an earlier peak and then a faster decline.

A more realistic scenario (and hence feasible assumption) is that there exists a time point $T>0$ (not necessarily equal to the peak time $t_V^*$) such that for  $0<\ain_1<\ain_2<1$:
$$\qquad V(t,\ain_1)\leq V(t,\ain_2),\mbox{ for } 0\leq t\leq T,$$ and the post-T reversal is not strong enough to compensate the early excess exposure, namely:
$$\int_T^\infty\left[G(\ain_1, V(t,\ain_1))-G(\ain_2, V(t,\ain_2))\right]_+\;dt \leq \int_0^T \left(G(\ain_2, V(t,\ain_2))-G(\ain_1, V(t,\ain_1))\right)\;dt.$$ Then, $\mathcal E(\ain_1)\leq \mathcal E(\ain_2)$. This condition is biologically meaningful: it says that the extra early viral amplification caused by larger $\ain$ dominates any later faster clearance.

In sensitivity form, $$\mathcal E'(\ain)= \int_0^\infty \frac{V(1+bV)+\ain \; \frac{d}{d\ain}V}{[1+(1-\ain)bV]^2}dt \geq 0,$$ 
i.e. the negative sensitivity of $V$ after the peak must not outweigh the positive early sensitivity and the direct increase of exposure caused by increasing $\ain$. For acute, single-wave infections, such as most respiratory viruses, where most target-cell loss occurs near the growth and peak phase, it is reasonable to assume that larger $\ain$ increases the dominant early viral burden and that any later reduction of $V$ is not large enough to compensate for it. The available numerical trajectories in Section \ref{SEC:num} all exhibit single-wave behaviour, which provides a plausible mechanism for the effective-exposure ordering in the simulated parameter regime. However, neither a general proof nor a counterexample for viral-load unimodality is presently available, so uniqueness of the viral peak is left as an open analytical question.

\end{remark}

\subsubsection{Possible biological/clinical interpretations of $\ain$}
The parameter $\ain\in[0,1]$ quantifies the relative contribution of instantaneous infection processes and their short-term dynamical history. For $\ain=1$, the model reduces to the classical Markovian within-host system, where transition rates depend exclusively on the current state. As $\ain$ decreases, derivative-dependent correction terms become increasingly important, introducing an effective inertial or memory effect into the infection dynamics. Biologically, this may represent finite intracellular processing times, delayed immune activation, receptor adaptation, or other mechanisms causing the host-virus interaction to respond not only to the present viral burden but also to its recent rate of change.
The classical model assumes that the host reacts perfectly to the instantaneous viral burden, whereas our model suggests that the response is damped by inertia, possibly correlating with aging,
immune exhaustion, chronic disease, immunosuppression, or other factors that suggest 
$\ain$ might be patient-specific. Thus, different patients (hosts) may have different values of a response parameter $\ain$. However, it would be too strong to say that $\ain$ is purely a host parameter. It affects the viral trajectory $V(t)$ and therefore the coupled host-virus process. A more precise interpretation is that $\ain$ is a host-virus dynamical responsiveness parameter. Its most interesting feature is that it can modify clinical-course quantities such as peak viral load and final tissue damage while leaving the local invasion threshold $R_0$ unchanged.

For example, suppose two patients have identical $b,p,d,m,H_0$. Then, $R_0$ is identical. Yet, if patient A has $\ain=1$ and patient B has $\ain=0.3$, then
the infection initially has the same invasion threshold in both patients, but 
patient B may exhibit, for example, a later viral peak, a lower peak, a longer infection duration, and
different final tissue damage. Thus, while $R_0$ is unchanged, the clinical course changes. One of the most interesting features of the model is precisely this separation. Our model separates two notions that are usually entangled: infectious potential and dynamical responsiveness.

Another important aspect is that our model separates infection establishment (via $R_0$ governed by $b,p,d,m,H_0$) and infection outcome (captured by $H_\infty$). From a modeling perspective, this separation is a strong argument that $\ain$ is not redundant with $b,p,d,m$. If $\ain$ changed $R_0$ and nothing else, it would merely be another infectivity parameter. The fact that it leaves $R_0$ unchanged while altering peak timing, peak height, duration, and final size suggests it governs a different biological mechanism.

\section{Numerical illustrations}\label{SEC:num}

In this section, we provide numerical solutions to the system \eqref{eq:H}--\eqref{eq:D}. The temporal illustrations include the analytically treated endpoint $\ain=0$, whereas all parameter grids, sensitivity surfaces, and continuation diagrams use the positive-order interval $0.1\leq\ain\leq1$. Throughout the section, we use the following set of model parameters that satisfy condition \eqref{eq:suff_condition} for all $\ain\in[0,1]$:\newpage
\begin{equation}\label{val:init}
	H_0 = 4\cdot10^{4},\qquad V_0=0,\qquad I_0 = 20, \qquad D_0 = 0,
\end{equation}
and
\begin{equation}\label{val:par}
	b=10^{-6},\qquad p=4\cdot10^{-2},\qquad d = 8\cdot10^{-3}, \qquad m = 2\cdot10^{-2}.
\end{equation}
Almost identical parameters were used also in \cite{Rost2021} as a set of parameter regime; here we only round up the initial number of healthy cells to 40.000 instead of 39.980 for better illustration purposes on the figures. In this section, we use these parameter values first to illustrate how the infection dynamics would theoretically change for $\ain\in\{0,0.1,0.2,0.3,0.4,0.5,0.6,0.7,0.8,0.9,1\}$ and note that the classical dynamics is recovered for $\ain=1$.

For this baseline parameter set, $R_0=10$ and the positive classical infected-subsystem eigenvalue is
\[
\mu_+=\frac{-(d+m)+\sqrt{(d-m)^2+4bpH_0}}{2}
\approx 0.0264475.
\]
The corresponding growing eigenvalue of the derivative-corrected system is
\[
\lambda_+(\ain)=\frac{\ain\mu_+}{1-(1-\ain)\mu_+},
\qquad
\frac{\dd\lambda_+}{\dd\ain}
=\frac{\mu_+(1-\mu_+)}{[1-(1-\ain)\mu_+]^2}>0.
\]
Thus increasing $\ain$ rigorously increases the dominant early exponential growth rate for the baseline parameters. %This spectral statement does not assert componentwise ordering of the full trajectories or, by itself, prove monotonicity of their peak times or peak magnitudes.

For the numerical implementation, we used Python and SciPy's \texttt{solve\_ivp} function with the LSODA method and relative and absolute tolerances $10^{-8}$ and $10^{-10}$, respectively. For $\ain>0$, the exact change of variable $\tau=\ain t$ was used, and the physical peak times were recovered as $t^\ast=\tau^\ast/\ain$. The peaks of $I$ and $V$ were located by downward zero crossings of $I'$ and $V'$, rather than by maxima on a sampled time grid. Starting from $t=16000$, the integration horizon was doubled whenever necessary until both peak events had been resolved and $I$ and $V$ had decayed to the prescribed numerical tail tolerance. The value reported as $H_\infty$ was accepted only after two consecutive tail checks and after further extension changed $H$ by less than the stated numerical tolerance. Conservation, the final-size identity, and positivity of the regularity determinant were also checked. At $\ain=0$, the nonsingular implicit system reduces to $X'=0$, so the trajectory is frozen and has no unique isolated peak.

\begin{figure}[!htbp]
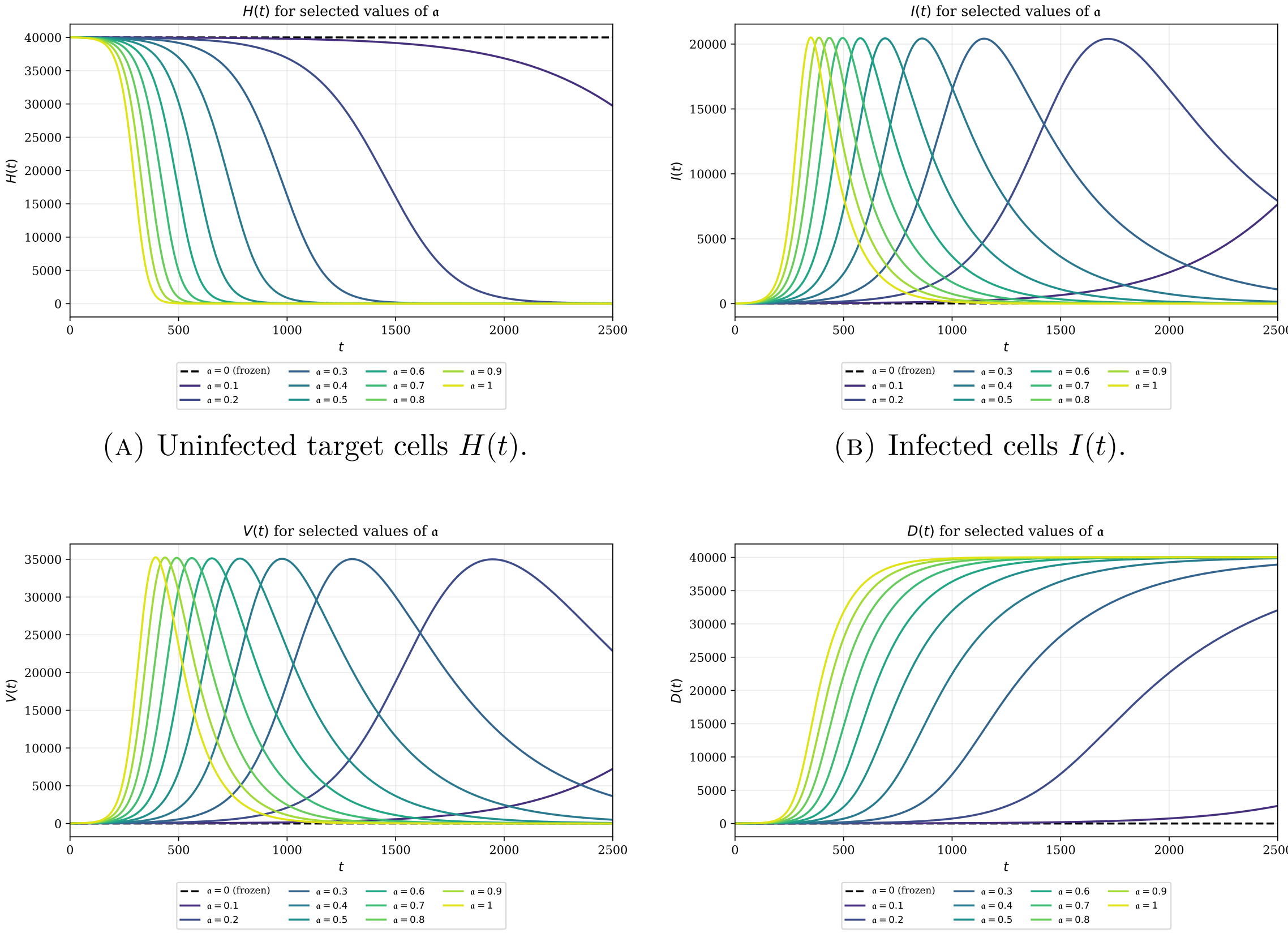

	\centering
	\subfloat[Uninfected target cells $H(t)$.\label{fig:H}]{\includegraphics[width=0.49\textwidth]{\certfig{Hsol}}}\hfill
	\subfloat[Infected cells $I(t)$.\label{fig:I}]{\includegraphics[width=0.49\textwidth]{\certfig{Isol}}}\\[0.5em]
	\subfloat[Viral load $V(t)$.\label{fig:V}]{\includegraphics[width=0.49\textwidth]{\certfig{Vsol}}}\hfill
	\subfloat[Damaged cells $D(t)$.\label{fig:D}]{\includegraphics[width=0.49\textwidth]{\certfig{Dsol}}}
	\caption{Temporal evolution of the four model compartments for the baseline parameter set and the indicated values of $\ain$.}%
	\label{fig:temporal-all}%
\end{figure}

\newpage
Figure~\ref{fig:temporal-all} illustrates the temporal evolution of all model compartments.

\FloatBarrier

For $0.1\leq\ain\leq1$, Table~\ref{tab:peaks} shows that the peak sizes decrease and the peak times are delayed as $\ain$ decreases. The computed final healthy-cell population remains strictly positive and increases as $\ain$ decreases for this parameter set.  At $\ain=0$, no isolated peak time is reported because the trajectory is constant; nevertheless, the positive-order peak times diverge to infinity as $\ain\downarrow0$. 

\begin{table}[!htbp]
	\centering
	\small
	\setlength{\tabcolsep}{4pt}
	\renewcommand{\arraystretch}{0.92}
	\begin{tabular}{|c|c|c|c|c|c|}
		\hline
		$\ain$ & $t_I^*$ & $I_{\max}$ & $t_V^*$ & $V_{\max}$ & $H_\infty$\\
		\hline
		0 & -- & 20.00 & -- & 0.00 & 40000.00 \\
		\hline
		0.1 & 3428.98 & 20386.83 & 3886.99 & 34964.61 & 2.21  \\
		\hline
		0.2 & 1717.92 & 20400.42 & 1946.82 & 34997.51 & 2.16 \\
		\hline
		0.3 & 1147.56 & 20414.05 & 1300.10 & 35030.50 & 2.11 \\
		\hline
		0.4 & 862.39 & 20427.70 & 976.73 & 35063.60 & 2.07  \\
		\hline
		0.5 & 691.28 & 20441.40 & 782.72 & 35096.81 & 2.02  \\
		\hline
		0.6 & 577.21 & 20455.12 & 653.37 & 35130.11 & 1.98 \\
		\hline
		0.7 & 495.73 & 20468.88 & 560.98 & 35163.52 & 1.93  \\
		\hline
		0.8 & 434.62 & 20482.67 & 491.69 & 35197.03 & 1.89  \\
		\hline
		0.9 & 387.09 & 20496.50 & 437.80 & 35230.65 & 1.85  \\
		\hline
		1 & 349.07 & 20510.36 & 394.68 & 35264.37 & 1.81  \\
		\hline
	\end{tabular}
	\caption{Peak times, peak heights, and  final healthy-cell population size (rounded to two decimals).}
	\label{tab:peaks}
\end{table}
\FloatBarrier

\begin{figure}[!htbp]
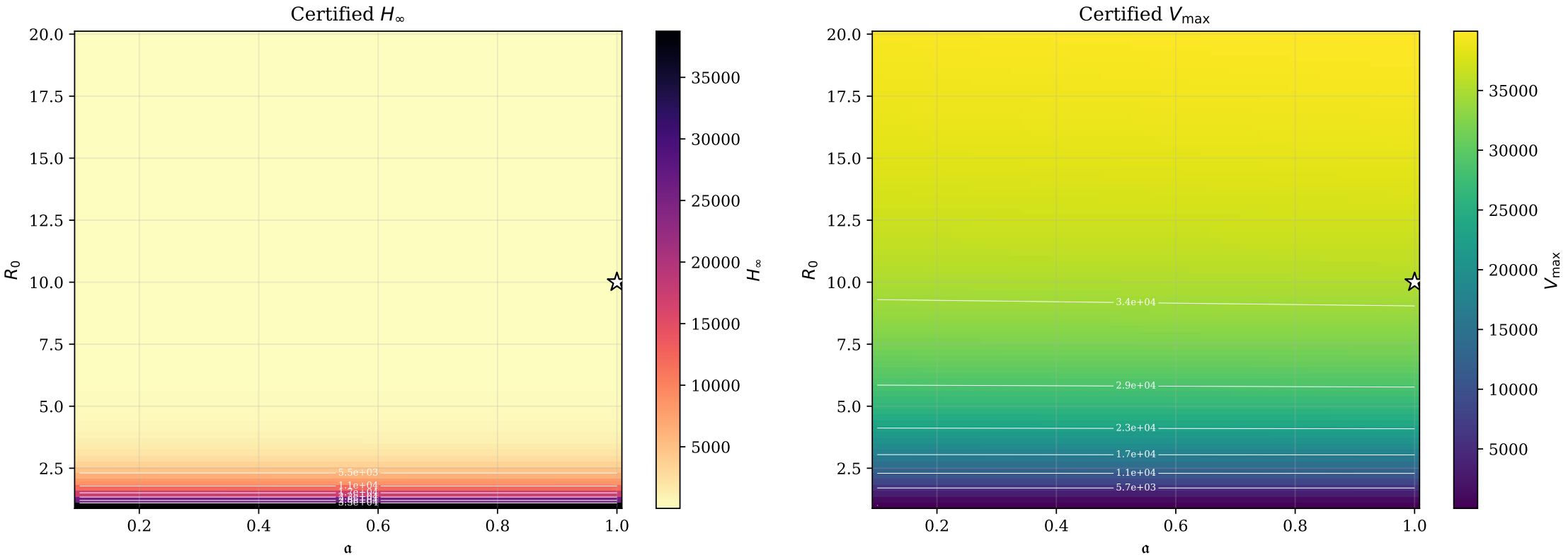
%
	\centering
	\subfloat[$H_\infty$ over $(\ain,R_0)$.]{{\includegraphics[width=0.48\textwidth]{\certfig{heatmapHinf}} }}%
	%\quad
	\subfloat[$V_{\max}$ over $(\ain,R_0)$.]{{\includegraphics[width=0.48\textwidth]{\certfig{heatmapVmax}} }}%
	\caption{Heat maps of the final healthy-cell population $H_\infty$ and maximal viral load $V_{\max}$ as functions of the derivative-correction parameter $\ain$ and the basic reproduction number $R_0$. The star indicates the baseline parameter set.}%
	\label{fig:heat1}%
\end{figure}

In addition to the temporal evolution simulations presented above, we perform a global sensitivity analysis to investigate how the qualitative behaviour of the model depends on the derivative-correction parameter $\ain$ and the basic reproduction number $R_0$. Rather than displaying individual solution trajectories, Figures~\ref{fig:heat1}--\ref{fig:heat2} summarize the model behaviour over a two-dimensional parameter space. The heat maps and the complementary numerical illustrations collected in Appendix~\ref{app:additional-numerics} were computed on $0.1\leq\ain\leq1$ and $1\leq R_0\leq20$, while keeping the remaining model parameters fixed at their baseline values. For each parameter pair, the integration horizon was extended adaptively until the event and asymptotic checks described above were satisfied; unresolved quantities would have been stored as missing and masked rather than replaced by the final integration time. All $4675$ grid runs passed these checks. Here and in the figures, a quantity is called \emph{certified} when the corresponding numerical run passed all applicable event-detection, tail-convergence, and consistency checks after adaptive extension of the integration interval. The terminology \emph{certified} indicates numerical validation within the prescribed tolerances.

Each point in the $(\ain,R_0)$-plane corresponds to one complete numerical simulation, from which biologically relevant quantities are extracted, including the final number of healthy cells $H_\infty$, the infected-cell peak time $t_I^\ast$, the maximal number of infected cells $I_{\max}$, and the maximal viral load $V_{\max}$. These quantities are represented by colour, producing heat maps in which regions of similar behaviour are readily identifiable. Thus, dark and light colours do not represent the solution itself, but rather the numerical value of the selected biological indicator at the corresponding parameter pair. At the threshold $R_0=1$, the infected-cell maximum occurs initially rather than at an interior turning point, so $t_I^\ast$ is undefined there and is masked in the peak-time plots. The baseline parameter set used throughout the numerical experiments corresponds to the classical case $\ain=1$ and $R_0=10$, which is indicated in each figure to facilitate comparison with the reference simulations presented earlier.

\begin{figure}[!htbp]
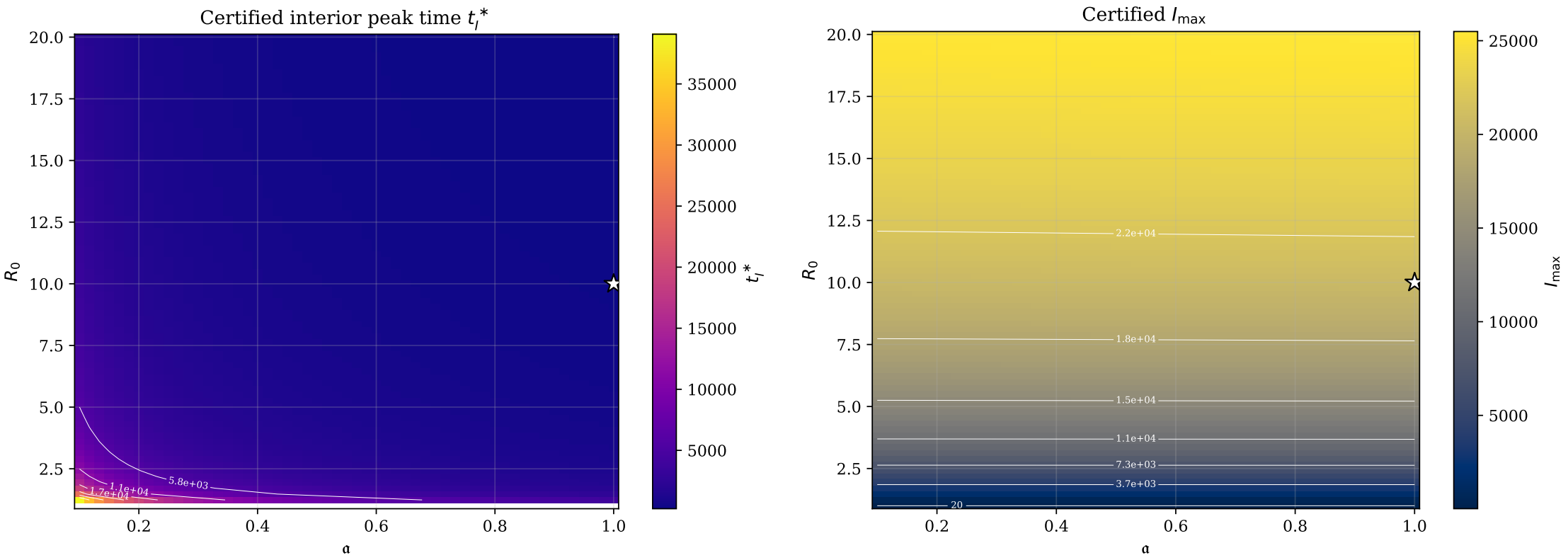
%
	\centering
	\subfloat[$t_I^\ast$ over $(\ain,R_0)$.]{{\includegraphics[width=0.48\textwidth]{\certfig{heatmapPeakI}} }}%
	%\quad
	\subfloat[$I_{\max}$ over $(\ain,R_0)$.]{{\includegraphics[width=0.48\textwidth]{\certfig{heatmapImax}} }}%
	\caption{Heat maps of the infected-cell peak time $t_I^\ast$ and peak size $I_{\max}$  over the $(\ain,R_0)$-parameter plane. The threshold row $R_0=1$ is masked in the peak-time panel because no interior infected-cell peak occurs there. The star indicates the baseline parameter set.}%
	\label{fig:heat2}%
\end{figure}
\FloatBarrier

The corresponding three-dimensional surfaces and finite-difference sensitivity-sign maps are given in Appendix~\ref{app:additional-numerics}. They provide alternative views of the same certified data and numerical support for the observed monotonicity patterns. The following parameter continuation diagrams (Figures~\ref{fig:cont1}--\ref{fig:cont2}) show directly how $H_\infty$, $I_{\max}$, $V_{\max}$, and $t_I^\ast$ vary with $\ain\in[0.1,1]$ for representative values of $R_0$. These plots are one-dimensional cross-sections of the two-dimensional parameter study and make the dependence on $\mathfrak{a}$ especially transparent. The endpoint $\ain=0$ is not inserted as a finite point in the peak-time continuation: its trajectory is frozen and has no isolated peak, whereas $t_I^\ast$ diverges to infinity along the positive-order continuation as $\ain\downarrow0$.

The almost horizontal appearance of some $H_\infty$ continuation curves is not caused by termination at a common round time. Each value is obtained after its own adaptive asymptotic check, and the logarithmic vertical scale is used to display the wide variation across $R_0$. In accordance with \eqref{eq:final_general}, $H_\infty$ depends on the complete effective viral-exposure trajectory, which changes only weakly with $\ain$ for some fixed values of $R_0$.

\begin{figure}[!htbp]
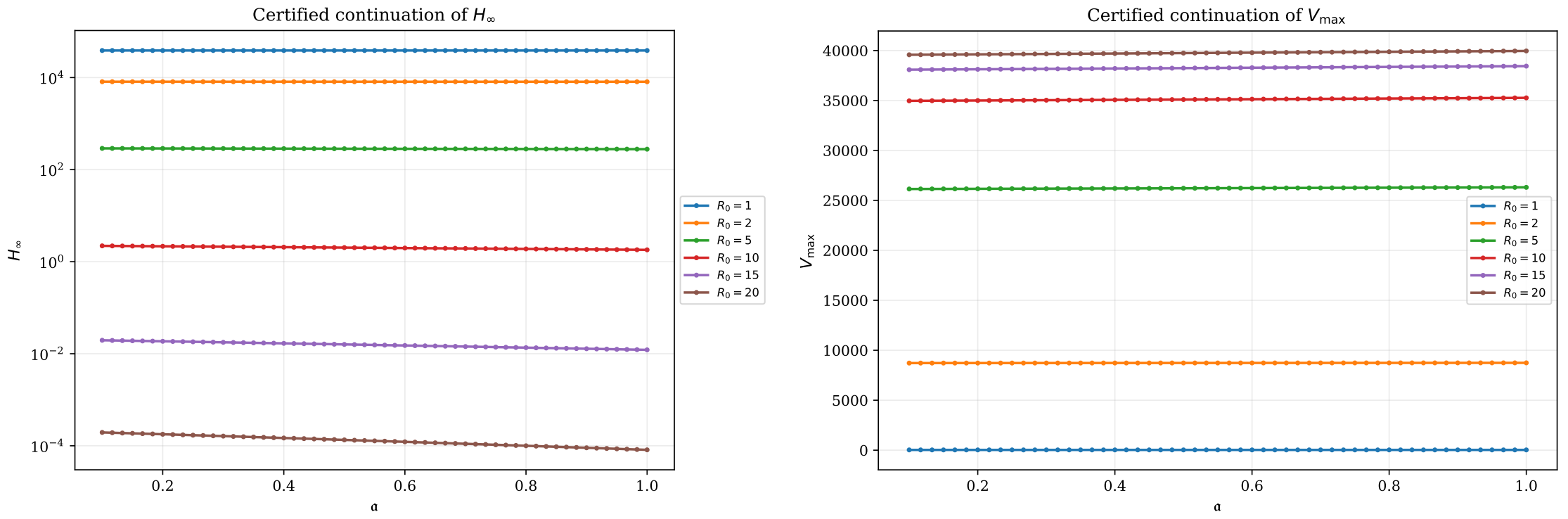
%
	\centering
	\subfloat[Continuation of $H_\infty$.]{{\includegraphics[width=0.48\textwidth]{\certfig{contHinf}} }}%
	%\quad
	\subfloat[Continuation of $V_{\max}$.]{{\includegraphics[width=0.48\textwidth]{\certfig{contVmax}} }}%
	\caption{Continuation of $H_\infty$ and $V_{\max}$ on $0.1\leq\ain\leq1$ for representative values of $R_0$. The $H_\infty$ panel uses a logarithmic vertical scale.}%
	\label{fig:cont1}%
\end{figure}

\begin{figure}[!htbp]
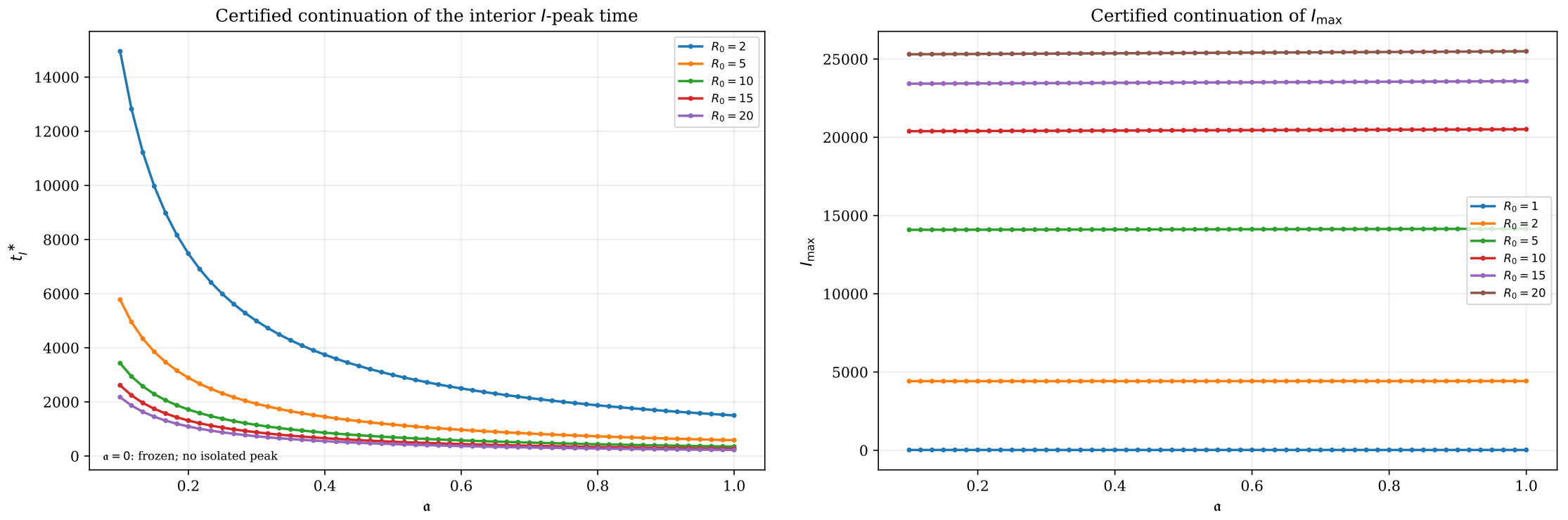
%
	\centering
	\subfloat[Continuation of $t_I^\ast$.]{{\includegraphics[width=0.48\textwidth]{\certfig{contPeakI}} }}%
	%\quad
	\subfloat[Continuation of $I_{\max}$.]{{\includegraphics[width=0.48\textwidth]{\certfig{contImax}} }}%
	\caption{Continuation of the infected-cell peak time $t_I^\ast$ and peak size $I_{\max}$ on $0.1\leq\ain\leq1$ for representative values of $R_0$. The $R_0=1$ peak-time curve is omitted because the infected-cell maximum is not an interior turning point; the positive-order peak time diverges as $\ain\downarrow0$.}%
	\label{fig:cont2}%
\end{figure}
\FloatBarrier

\begin{figure}[!htbp]
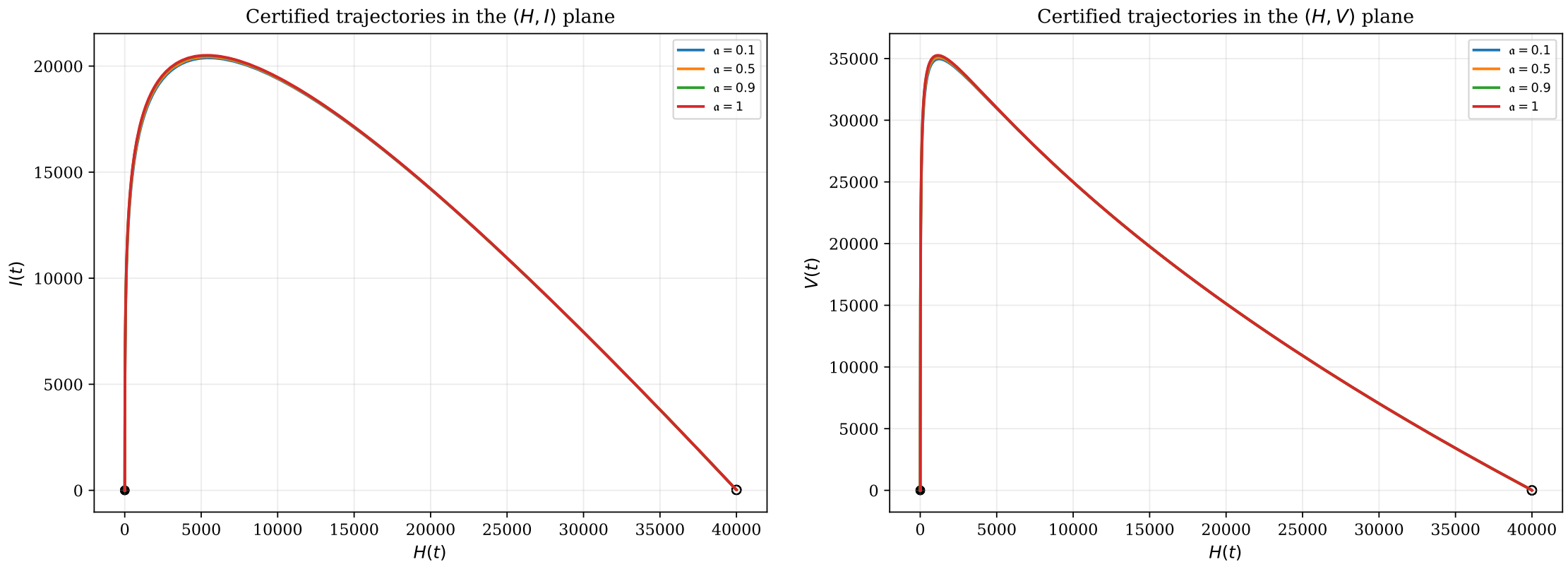
%
	\centering
	\subfloat[$(H,I)$ phase plane.]{{\includegraphics[width=0.48\textwidth]{\certfig{phaseHI}} }}%
	%\quad
	\subfloat[$(H,V)$ phase plane.]{{\includegraphics[width=0.48\textwidth]{\certfig{phaseHV}} }}%
	\caption{Projected phase trajectories in the $(H,I)$ and $(H,V)$ planes for selected values of $\ain$, showing the movement from the initial state toward the long-term equilibrium.}%
	\label{fig:phase-diagrams}%
\end{figure}

\begin{figure}[!htbp]%
	\centering
	\includegraphics[width=0.8\textwidth]{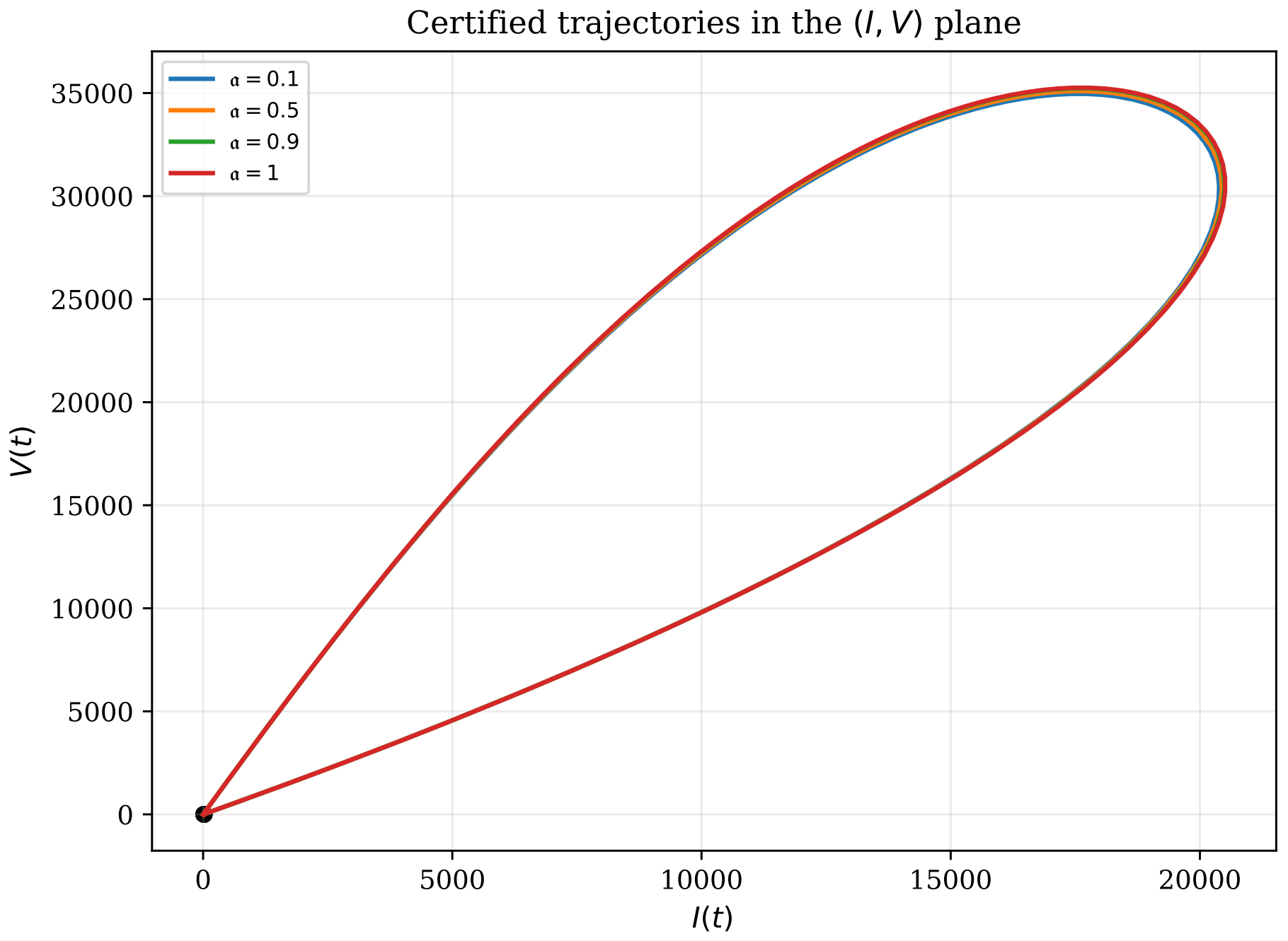}
	\caption{Projected phase trajectory in the $(I,V)$ plane for selected values of $\ain$, showing the relation between infected cells and viral load during the transient infection.}%
	\label{fig:phase-IV}%
\end{figure}

The phase diagrams in Figures~\ref{fig:phase-diagrams}--\ref{fig:phase-IV} depict the evolution of pairs of state variables, such as $(H,I)$, $(H,V)$, and $(I,V)$, independently of time. Rather than showing the temporal evolution of each compartment separately, these diagrams represent the solution trajectory in the corresponding phase plane and illustrate the interactions between biological populations during infection. The geometry of the trajectories reveals characteristic features of the disease dynamics, including the depletion of healthy cells, the transient growth and subsequent decline of infected cells and viral particles, and convergence toward the long-term equilibrium.

Comparing phase trajectories for different values of $\mathfrak{a}$ provides an intuitive visualization of how the derivative-corrected dynamics modify the overall infection pathway. Changes in the curvature, excursion, and terminal location of the trajectories quantify the influence of the derivative-correction mechanism on infection progression, complementing the information obtained from the sensitivity analyses.

From Figures~\ref{fig:phase-diagrams}--\ref{fig:phase-IV} $\ain$ seems to act predominantly as a time-scale parameter: it strongly changes how quickly the infection trajectory is traversed while producing only modest changes in its phase-space geometry and final target-cell level. However, this approximate time-reparametrization property is numerical rather than exact or universal, and it relates to the baseline parameters chosen in \eqref{val:init}--\eqref{val:par}.
\FloatBarrier

Additional illustrations of the complete cell-population structure, the equilibrium manifold, and the dependence on initial conditions are collected in Appendix~\ref{app:additional-numerics}. 

\section{Conclusion}

In this paper, we developed a within-host infection system that can be interpreted as a convex mixture between the classical model and a derivative-corrected model. The classical model assumes that the rates respond instantaneously to the current state: $H'=-bVH$, $I'=bVH-dI$, $V'=pI-mV$, while our derivative-corrected dynamical system model instead states that the system evolves according to a mixture of
\begin{itemize}
\item the instantaneous rate (weighted by $\ain$),
\item the change of the rate itself (weighted by $1-\ain$).
\end{itemize}
Thus, the loss of healthy cells depends not only on the current exposure $VH$, but also on how rapidly the exposure is changing, as captured by $(VH)'$.
Biologically, this could represent gradual receptor internalization, intracellular signaling cascades, finite activation times, physiological inertia or other factors reflecting the fact that the host does not respond instantaneously.

It is also worth noting that classical within-host models sometimes include explicit delays, e.g. $I'(t)=bV(t-\tau)H(t-\tau)-dI(t),$ where $\tau$ is the eclipse phase. Our model contains derivative corrections such as $f'(t)=(VH)'$, which are formally very similar to first-order approximations of delayed terms:$$f(t-\tau)=f(t)-\tau f'(t)+O(\tau^2),$$ whereas our model contains a positive multiple of $f'(t)$.
%Thus, $1-\ain$ may biologically measure the strength of an effective intracellular delay, thereby improving the classical model, which neglects the time required for viral entry, replication, and immune recognition.
Although both mechanisms can alter transient time scales and describe non-instantaneous biological responses, our model is not equivalent to a delayed within-host model or a model that would depend on future states, the similarity is only formal, and $\ain$ might be related to measure the strength of an effective intracellular delay or other biological mechanisms in comparison to the classical ODE model which neglects the time required for viral entry, replication, and immune recognition.

We have proved that the model has a unique positive solution, derived final size estimates, analyzed peak times and peak values that are influenced by the parameter $\ain$, and showed that the basic reproduction number does not depend on $\ain$. This feature arises because near the disease-free equilibrium the implicit correction affects both production and removal pathways in the same way, and these corrections cancel out in the next-generation matrix. Other features, such as the viral peak, infection peak, and final size of remaining healthy cells, are strongly dependent on $\ain$.

Finally, we note that this paper does not address fitting the model to real data, estimating the parameter values $b,p,d,m$, or estimating $\ain$. The values stated in \eqref{val:init}--\eqref{val:par} were used for illustrative purposes only. Fitting the model to available viral-load data and benchmarking it against other within-host dynamical models will be addressed in a forthcoming research \cite{paperNO2}.

\clearpage
\appendix
\section{Additional numerical illustrations}\label{app:additional-numerics}

This appendix collects complementary numerical visualizations based on the same certified computations described in Section~\ref{SEC:num}. 

\begin{figure}[!htbp]
	\centering
	\includegraphics[width=0.82\textwidth]{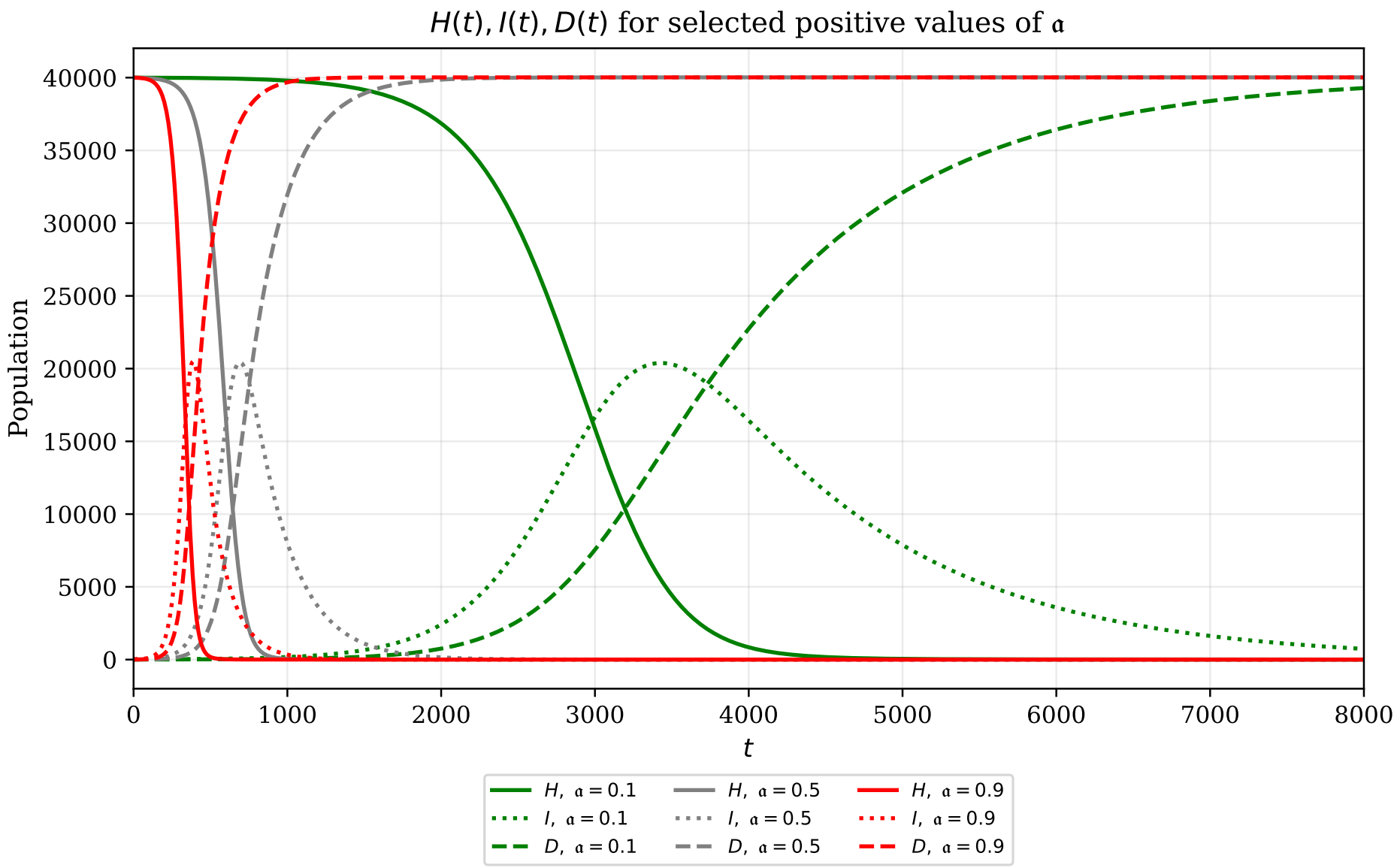}
	\caption{Temporal evolution of the full cell-population structure for selected positive values of $\ain$.}%
	\label{fig:population}%
\end{figure}

Figure~\ref{fig:population} displays the temporal evolution of the full cell-population structure for selected positive values of $\ain$ and complements the compartment-wise plots in Figure~\ref{fig:temporal-all}.

\FloatBarrier

\begin{figure}[!htbp]
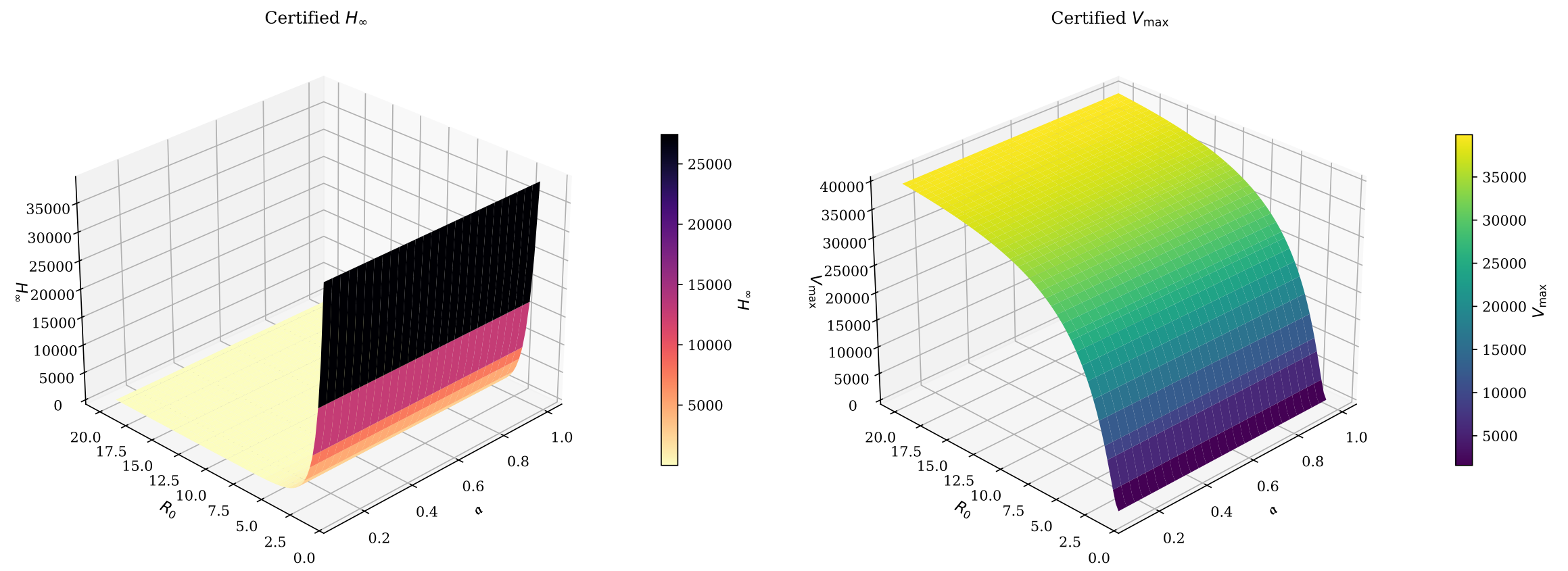

	\centering
	\subfloat[Surface for $H_\infty$.]{\includegraphics[width=0.49\textwidth]{\certfig{surfHinf}}}\hfill
	\subfloat[Surface for $V_{\max}$.]{\includegraphics[width=0.49\textwidth]{\certfig{surfVmax}}}
	\caption{Three-dimensional sensitivity surfaces of $H_\infty$ and $V_{\max}$ over the $(\ain,R_0)$-parameter plane.}%
	\label{fig:surf1}%
\end{figure}

\begin{figure}[!htbp]
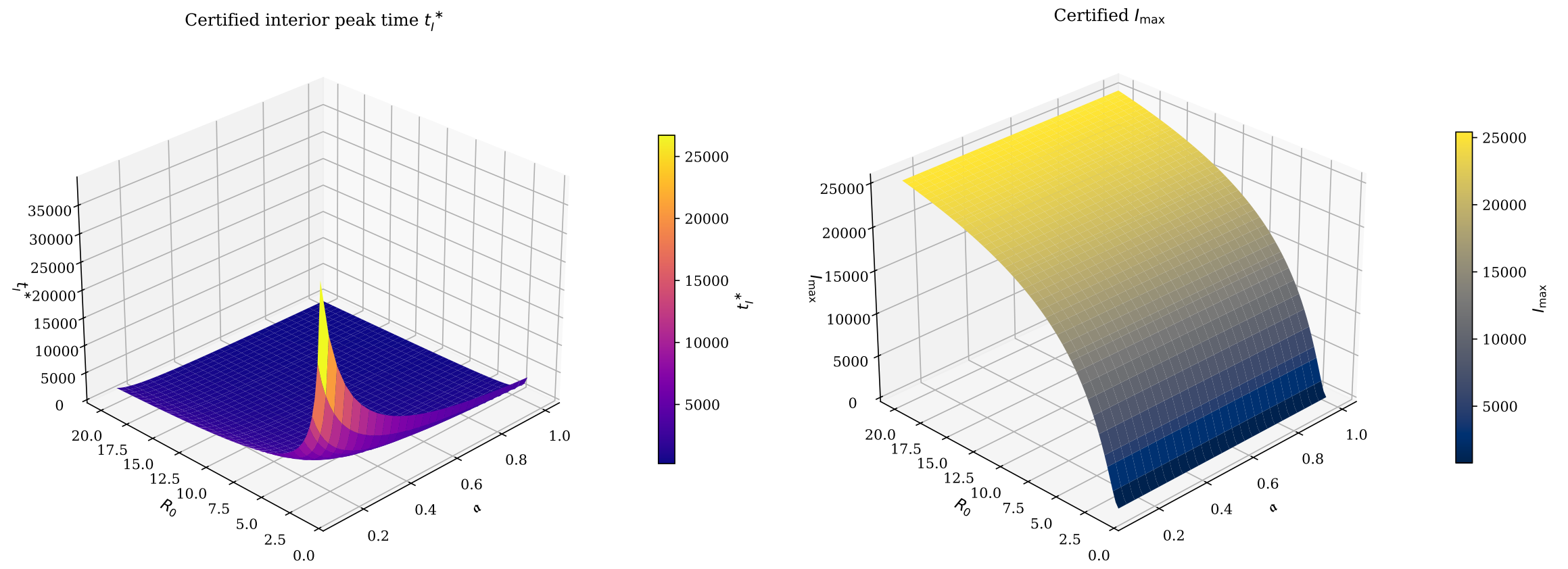

	\centering
	\subfloat[Surface for $t_I^\ast$.]{\includegraphics[width=0.49\textwidth]{\certfig{surfPeakI}}}\hfill
	\subfloat[Surface for $I_{\max}$.]{\includegraphics[width=0.49\textwidth]{\certfig{surfImax}}}
	\caption{Three-dimensional sensitivity surfaces of the infected-cell peak time $t_I^\ast$ and peak size $I_{\max}$ over the $(\ain,R_0)$-parameter plane. The undefined threshold values of $t_I^\ast$ at $R_0=1$ are masked.}%
	\label{fig:surf2}%
\end{figure}

The three-dimensional sensitivity surfaces in Figures~\ref{fig:surf1}--\ref{fig:surf2} provide an alternative visualization of the heat-map data in Figures~\ref{fig:heat1}--\ref{fig:heat2}. In each panel, the horizontal axis represents $\ain\in[0.1,1]$ and the other horizontal axis represents $R_0\in[1,20]$. The surfaces facilitate visual identification of monotonic trends, transition regions, and parameter interactions.

\FloatBarrier

\begin{figure}[!htbp]
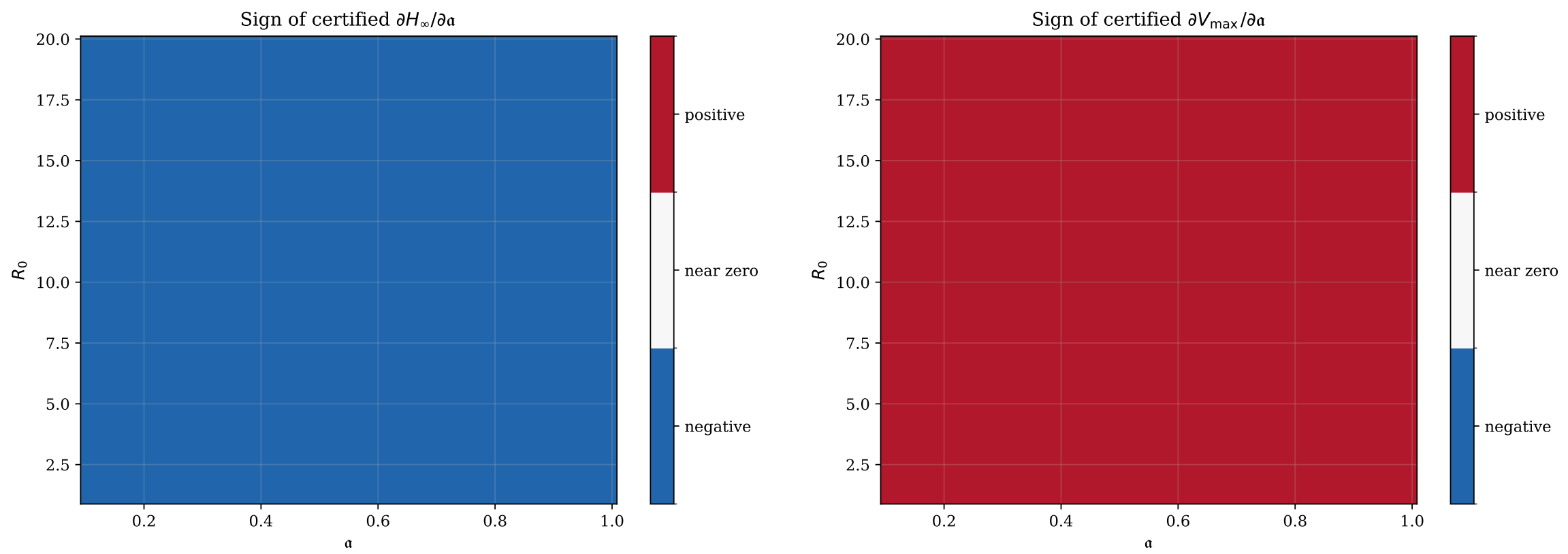

	\centering
	\subfloat[Sign of $\partial H_\infty/\partial\ain$.]{\includegraphics[width=0.49\textwidth]{\certfig{signsensHinf}}}\hfill
	\subfloat[Sign of $\partial V_{\max}/\partial\ain$.]{\includegraphics[width=0.49\textwidth]{\certfig{signsensVmax}}}
	\caption{Signs of the numerical sensitivities of the certified values of $H_\infty$ and $V_{\max}$ with respect to $\ain$ on $0.1\leq\ain\leq1$.}%
	\label{fig:sign1}%
\end{figure}

\begin{figure}[!htbp]
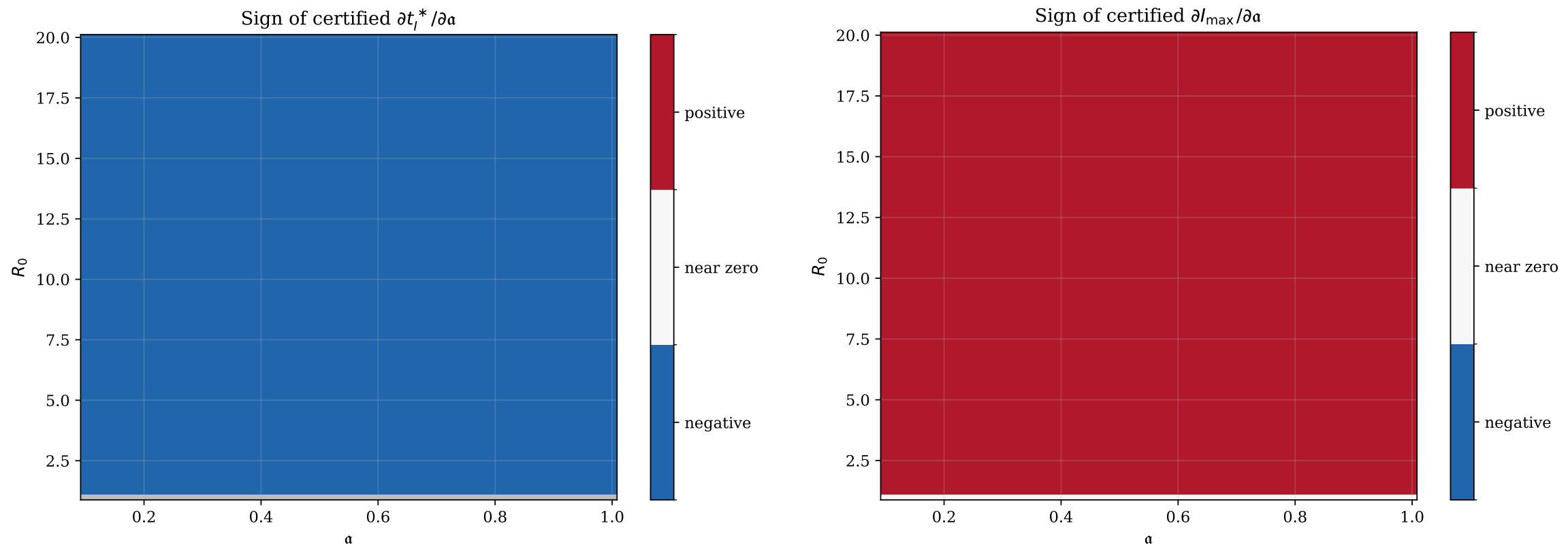

	\centering
	\subfloat[Sign of $\partial t_I^\ast/\partial\ain$.]{\includegraphics[width=0.49\textwidth]{\certfig{signsensPeakI}}}\hfill
	\subfloat[Sign of $\partial I_{\max}/\partial\ain$.]{\includegraphics[width=0.49\textwidth]{\certfig{signsensImax}}}
	\caption{Signs of the numerical sensitivities of the certified infected-cell peak time $t_I^\ast$ and peak size $I_{\max}$ with respect to $\ain$ on $0.1\leq\ain\leq1$. Undefined threshold values are masked.}%
	\label{fig:sign2}%
\end{figure}

We also computed the local sensitivities $\frac{d}{d\ain}H_\infty$, $\frac{d}{d\ain}V_{\max}$, $\frac{d}{d\ain}t_I^\ast$, and $\frac{d}{d\ain}I_{\max}$ by second-order finite differences using only certified finite values. Undefined regions are masked in Figures~\ref{fig:sign1}--\ref{fig:sign2}. These numerical derivatives provide numerical support for the monotonicity patterns under the parameter regime considered.

\FloatBarrier

\begin{figure}[!htbp]
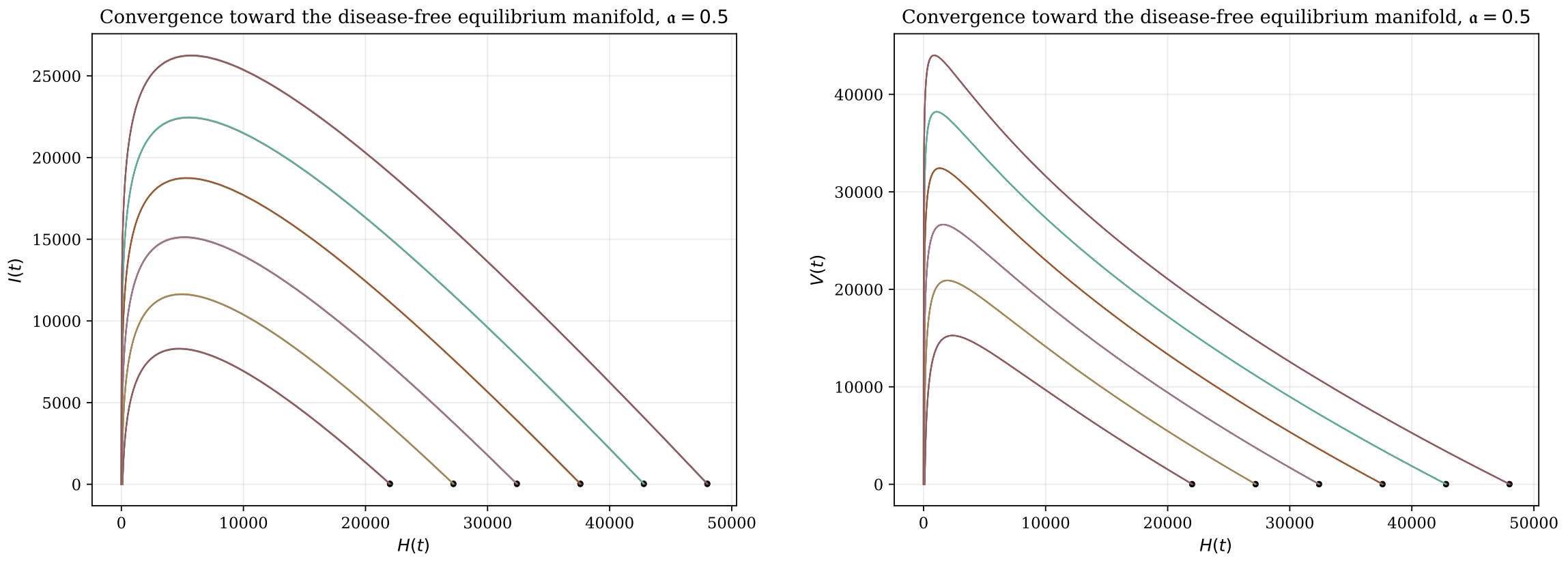

	\centering
	\subfloat[$(H,I)$ plane.]{\includegraphics[width=0.49\textwidth]{\certfig{basinHI05}}}\hfill
	\subfloat[$(H,V)$ plane.]{\includegraphics[width=0.49\textwidth]{\certfig{basinHV05}}}
	\caption{Convergence toward disease-free equilibrium states on the manifold for $\ain=0.5$ from several initial states.}%
	\label{fig:basin-attractors}%
\end{figure}

The disease-free equilibria form a continuum rather than a single attracting equilibrium. Figure~\ref{fig:basin-attractors} illustrates convergence toward this equilibrium manifold from several initial states. In the $(H,I)$ projection we vary $(H_0,I_0)$, while in the $(H,V)$ projection we vary $(H_0,V_0)$. The trajectories generally approach different limiting points $(H_\infty,0)$, illustrating the dependence of the terminal state on $(H_0,I_0,V_0,D_0)$.

\FloatBarrier

\begin{figure}[!htbp]
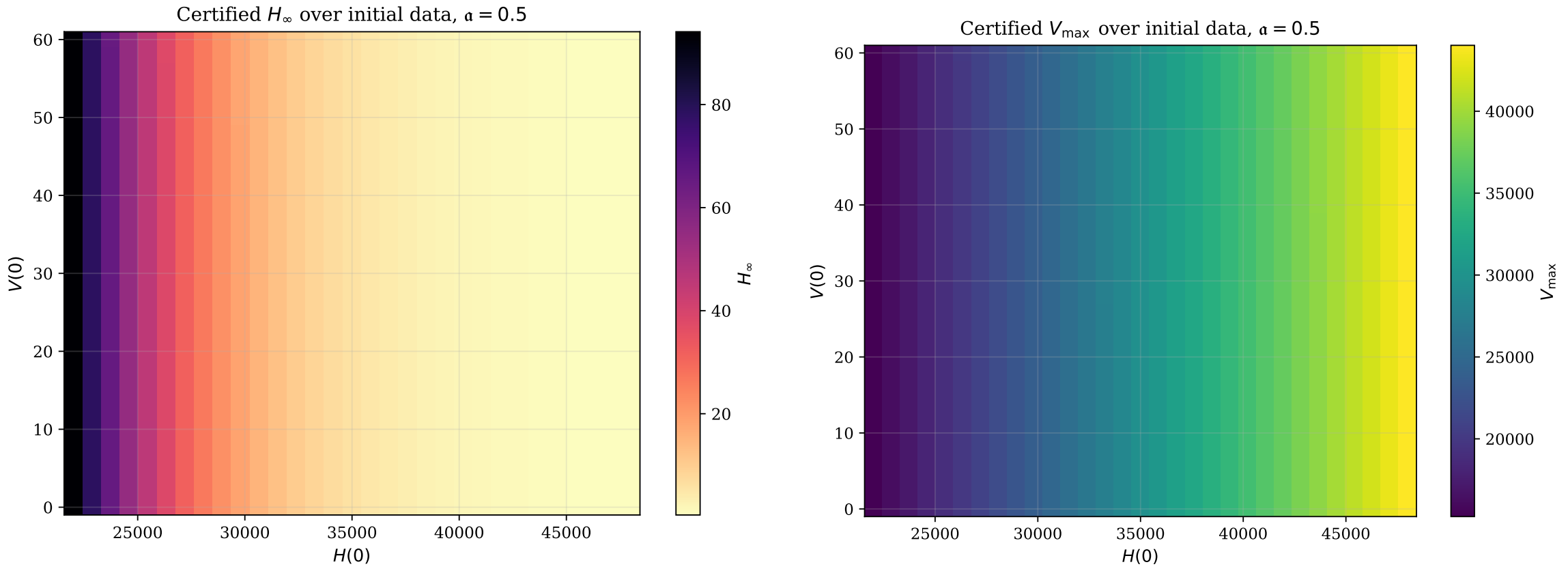

	\centering
	\subfloat[Final population $H_\infty$.]{\includegraphics[width=0.49\textwidth]{\certfig{initHinf}}}\hfill
	\subfloat[Maximum viral load $V_{\max}$.]{\includegraphics[width=0.49\textwidth]{\certfig{initVmax}}}
	\caption{Dependence of $H_\infty$ and $V_{\max}$ on the initial uninfected target-cell abundance $H_0$ and initial viral load $V_0$ for $\ain=0.5$.}%
	\label{fig:init1}%
\end{figure}

\begin{figure}[!htbp]
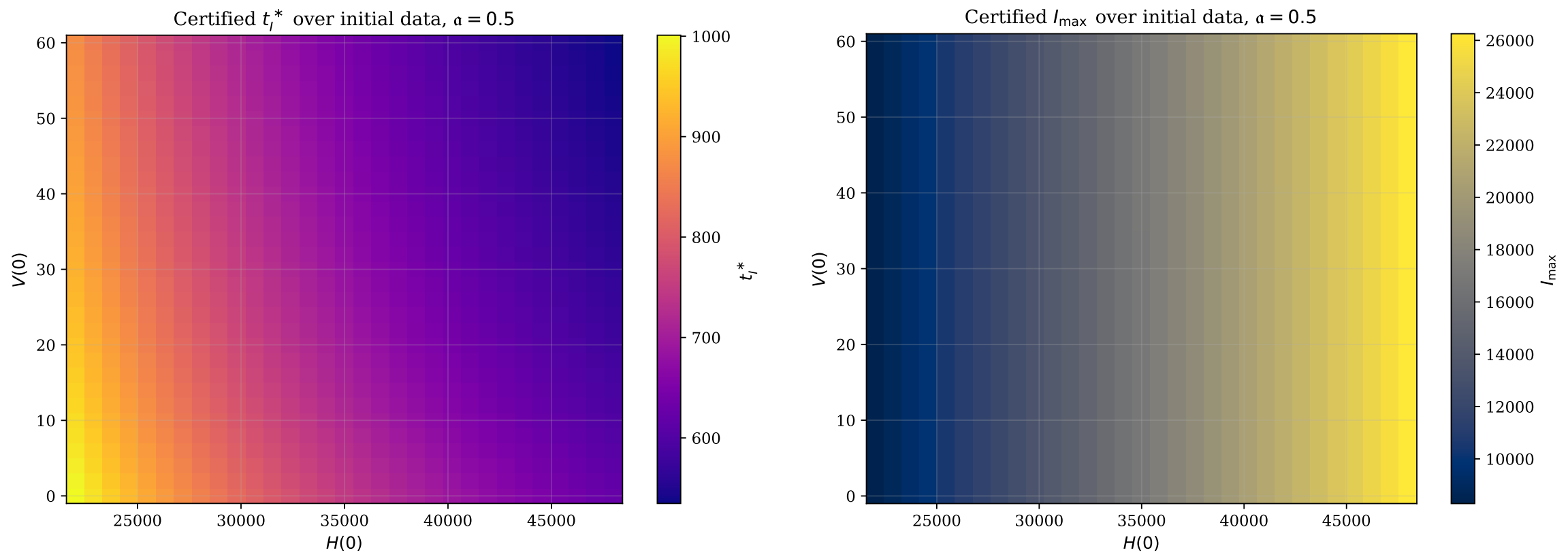

	\centering
	\subfloat[Peak time $t_I^\ast$.]{\includegraphics[width=0.49\textwidth]{\certfig{initPeakI}}}\hfill
	\subfloat[Peak size $I_{\max}$.]{\includegraphics[width=0.49\textwidth]{\certfig{initImax}}}
	\caption{Dependence of the infected-cell peak time $t_I^\ast$ and peak size $I_{\max}$ on the initial uninfected target-cell abundance $H_0$ and initial viral load $V_0$ for $\ain=0.5$.}%
	\label{fig:init2}%
\end{figure}

Although $R_0$ depends on $H_0$ and is independent of $V_0$, the initial viral load still affects the infection dynamics. The value of $R_0$ determines the local invasion threshold, whereas $V_0$ controls the initial viral exposure and shapes the transient response. Consequently, increasing $V_0$ raises the cumulative viral exposure and reduces the final uninfected target-cell population. Figures~\ref{fig:init1}--\ref{fig:init2} show the dependence of $H_\infty$, $V_{\max}$, $t_I^\ast$, and $I_{\max}$ on $(H_0,V_0)$.

\FloatBarrier

\begin{acknowledgements}
The research and all authors were financially supported by the bilateral project Hungary-Serbia \emph{"Mathematical modeling of within-host dynamics and pharmacological aspects of SARS-CoV-2 infection"} running through 2021\slash23. DS and DR\'C gratefully acknowledge the financial support of the Ministry of Science, Technological Development and Innovation of the Republic of Serbia (Grants No. 451-03-33\slash2026-03\slash200125 and 451-03-34\slash2026-03\slash200125). NJ and GR gratefully acknowledge the support by 2019-2.1.11-T\'ET-2020-00231, 2024-1.2.3-HU-RIZONT-2024-00034 and TKP2021-EGA-05. NJ also acknowledges support by Project No. 152347, funded by the Ministry of Science and Technology of Hungary through the National Research, Development and Innovation Fund under the STARTING 25 funding scheme.
\end{acknowledgements}

%\bibliographystyle{kjmath}
%\bibliography{bibliography-ODE}

\begin{thebibliography}{10}
	
	\bibitem{AlzahraniElDessokyBaleanu2021}
	E.~Alzahrani, M.~M. El-Dessoky and D.~Baleanu, \textit{Modeling the dynamics of
		the novel coronavirus using {C}aputo--{F}abrizio derivative}, Alex. Eng. J.
	\textbf{60} (2021), 4651--4662,
	\href{https://doi.org/10.1016/j.aej.2021.03.028}{doi:
		\nolinkurl{10.1016/j.aej.2021.03.028}}.
	
	\bibitem{AtanackovicJanevPilipovic2019}
	T.~M. Atanackovi{\'c}, M.~Janev and S.~Pilipovi{\'c}, \textit{Wave equation in
		fractional {Z}ener-type viscoelastic media involving {C}aputo--{F}abrizio
		fractional derivatives}, Meccanica \textbf{54} (2019), 155--167,
	\href{https://doi.org/10.1007/s11012-018-0920-5}{doi:
		\nolinkurl{10.1007/s11012-018-0920-5}}.
	
	\bibitem{AtanackovicPilipovicZorica2018}
	T.~M. Atanackovi{\'c}, S.~Pilipovi{\'c} and D.~Zorica, \textit{Properties of
		the {C}aputo--{F}abrizio fractional derivative and its distributional
		settings}, Fract. Calc. Appl. Anal. \textbf{21} (2018), 29--44,
	\href{https://doi.org/10.1515/fca-2018-0003}{doi:
		\nolinkurl{10.1515/fca-2018-0003}}.
	
	\bibitem{Baccam2006}
	P.~Baccam, C.~Beauchemin, C.~A. Macken, F.~G. Hayden and A.~S. Perelson,
	\textit{Kinetics of influenza {A} virus infection in humans}, J. Virol.
	\textbf{80} (2006), 7590--7599,
	\href{https://doi.org/10.1128/JVI.01623-05}{doi:
		\nolinkurl{10.1128/JVI.01623-05}}.
	
	\bibitem{Beauchemin2011}
	C.~A.~A. Beauchemin and A.~Handel, \textit{A review of mathematical models of
		influenza {A} infections within a host or cell culture: Lessons learned and
		challenges ahead}, BMC Public Health \textbf{11} (2011), S7,
	\href{https://doi.org/10.1186/1471-2458-11-S1-S7}{doi:
		\nolinkurl{10.1186/1471-2458-11-S1-S7}}.
	
	\bibitem{CaputoFabrizio2015}
	M.~Caputo and M.~Fabrizio, \textit{A new definition of fractional derivative
		without singular kernel}, Prog. Fract. Differ. Appl. \textbf{1} (2015),
	73--85,
	\href{https://doi.org/10.12785/PFDA/010201}{doi:
		\nolinkurl{10.12785/PFDA/010201}}.
	
	\bibitem{CaputoFabrizio2016Applications}
	M.~Caputo and M.~Fabrizio, \textit{Applications of new time and spatial
		fractional derivatives with exponential kernels}, Prog. Fract. Differ. Appl.
	\textbf{2} (2016), 1--11,
	\href{https://doi.org/10.18576/pfda/020101}{doi:
		\nolinkurl{10.18576/pfda/020101}}.
	
	\bibitem{Chen2021}
	Y.~Chen, J.~Cheng, Y.~Jiang and K.~Liu, \textit{Review of fractional epidemic
		models}, Appl. Math. Model. \textbf{97} (2021), 281--307,
	\href{https://doi.org/10.1016/j.apm.2021.03.044}{doi:
		\nolinkurl{10.1016/j.apm.2021.03.044}}.
	
	\bibitem{CiupeHeffernan2017}
	S.~M. Ciupe and J.~M. Heffernan,
	\textit{In-host modeling},
	Infect. Dis. Model. \textbf{2} (2017), 188--202,
	\href{https://doi.org/10.1016/j.idm.2017.04.002}{doi:
		\nolinkurl{10.1016/j.idm.2017.04.002}}.
	
	\bibitem{Ciupe2007}
	S.~M. Ciupe, R.~M. Ribeiro, P.~W. Nelson and A.~S. Perelson,
	\textit{Modeling the mechanisms of acute hepatitis B virus infection},
	J. Theor. Biol. \textbf{247} (2007), 23--35,
	\href{https://doi.org/10.1016/j.jtbi.2007.02.017}{doi:
		\nolinkurl{10.1016/j.jtbi.2007.02.017}}.
	
	\bibitem{CorreaEscuderoGomezAguilarLopezLopezAlvaradoMartinezBaleanu2022}
	I.~L. Correa-Escudero, J.~F. G{\'o}mez-Aguilar, M.~G. L{\'o}pez-L{\'o}pez,
	V.~M. Alvarado-Mart{\'i}nez and D.~Baleanu, \textit{Correcting dimensional
		mismatch in fractional models with power, exponential and proportional kernel:
		Application to electrical systems}, Results Phys. \textbf{40} (2022), 105867,
	\href{https://doi.org/10.1016/j.rinp.2022.105867}{doi:
		\nolinkurl{10.1016/j.rinp.2022.105867}}.
	
	\bibitem{Diekmann1990}
	O.~Diekmann, J.~A.~P. Heesterbeek and J.~A.~J. Metz, \textit{On the definition
		and the computation of the basic reproduction ratio {$R_0$} in models for
		infectious diseases in heterogeneous populations}, J. Math. Biol. \textbf{28}
	(1990), 365--382,
	\href{https://doi.org/10.1007/BF00178324}{doi:
		\nolinkurl{10.1007/BF00178324}}.
	
	\bibitem{DingYe2009}
	Y.~Ding and H.~Ye, \textit{A fractional-order differential equation model of
		{HIV} infection of {CD4+} t-cells}, Math. Comput. Model. \textbf{50} (2009),
	386--392,
	\href{https://doi.org/10.1016/j.mcm.2009.04.019}{doi:
		\nolinkurl{10.1016/j.mcm.2009.04.019}}.
	
	\bibitem{EftimieGillardCantrell2016}
	R.~Eftimie, J.~J. Gillard and D.~A. Cantrell, \textit{Mathematical models for
		immunology: Current state of the art and future research directions}, Bull.
	Math. Biol. \textbf{78} (2016), 2091--2134,
	\href{https://doi.org/10.1007/s11538-016-0214-9}{doi:
		\nolinkurl{10.1007/s11538-016-0214-9}}.
	
	\bibitem{Eubank2004}
	S.~Eubank, H.~Guclu, V.~S.~A. Kumar, M.~V. Marathe, A.~Srinivasan,
	Z.~Toroczkai and N.~Wang, \textit{Modelling disease outbreaks in realistic
		urban social networks}, Nature \textbf{429} (2004), 180--184,
	\href{https://doi.org/10.1038/nature02541}{doi:
		\nolinkurl{10.1038/nature02541}}.
	
	\bibitem{Gergely}
	S.~A. Gourley, G.~R\"ost and H.~R. Thieme,
	\textit{Uniform persistence in a model for bluetongue dynamics},
	SIAM J. Math. Anal. \textbf{46} (2014), 1160--1184,
	\href{https://doi.org/10.1137/120878197}{doi:
		\nolinkurl{10.1137/120878197}}.
	
	\bibitem{HarroucheMomaniHasanAlSmadi2021}
	N.~Harrouche, S.~Momani, S.~Hasan and M.~Al-Smadi, \textit{Computational
		algorithm for solving drug pharmacokinetic model under uncertainty with
		nonsingular kernel type {C}aputo--{F}abrizio fractional derivative}, Alex.
	Eng. J. \textbf{60} (2021), 4347--4362,
	\href{https://doi.org/10.1016/j.aej.2021.03.016}{doi:
		\nolinkurl{10.1016/j.aej.2021.03.016}}.
	
	\bibitem{Hethcote2000}
	H.~W. Hethcote, \textit{The mathematics of infectious diseases}, SIAM Rev.
	\textbf{42} (2000), 599--653,
	\href{https://doi.org/10.1137/S0036144500371907}{doi:
		\nolinkurl{10.1137/S0036144500371907}}.
	
	\bibitem{paperNO2}
	N.~Juh{\'a}sz, G.~R\"ost, F.~A. Bartha,  D.~Nagy, B.~Das, D.~Sele{\v{s}}i and
	D.~Rajter-\'Ciri\'c  
	\textit{Fitting and selection of within-host models for
		\textup{SARS-CoV-2} infection}, preprint.
	
	\bibitem{KeelingEames2005}
	M.~J. Keeling and K.~T.~D. Eames, \textit{Networks and epidemic models},
	J. R. Soc. Interface \textbf{2} (2005), 295--307,
	\href{https://doi.org/10.1098/rsif.2005.0051}{doi:
		\nolinkurl{10.1098/rsif.2005.0051}}.
	
	\bibitem{Kilbas2006}
	A.~A. Kilbas, H.~M. Srivastava and J.~J. Trujillo, \textit{Theory and
		Applications of Fractional Differential Equations}, Elsevier, 2006,
	\href{https://doi.org/10.1016/S0304-0208(06)80001-0}{doi:
		\nolinkurl{10.1016/S0304-0208(06)80001-0}}.
	
	\bibitem{KunkelMehrmann2006}
	P.~Kunkel and V.~Mehrmann, \textit{Differential-Algebraic Equations: Analysis
		and Numerical Solution}, European Mathematical Society, 2006,
	\href{https://doi.org/10.4171/017}{doi: \nolinkurl{10.4171/017}}.
	
	\bibitem{LosadaNieto2015}
	J.~Losada and J.~J. Nieto, \textit{Properties of a new fractional derivative
		without singular kernel}, Prog. Fract. Differ. Appl. \textbf{1} (2015),
	87--92,
	\href{https://doi.org/10.12785/PFDA/010202}{doi:
		\nolinkurl{10.12785/PFDA/010202}}.
	
	\bibitem{Luchko2020GeneralDerivative}
	Y.~Luchko, \textit{The general fractional derivative and related fractional
		differential equations}, Mathematics \textbf{8} (2020), 2115,
	\href{https://doi.org/10.3390/math8122115}{doi:
		\nolinkurl{10.3390/math8122115}}.
	
	\bibitem{Rost2021}
	S.~Marzban, R.~Han, N.~Juh{\'a}sz and G.~R{\"o}st, \textit{A hybrid
		{PDE}--{ABM} model for viral dynamics with application to {SARS-CoV-2} and
		influenza}, R. Soc. Open Sci. \textbf{8} (2021), 210787,
	\href{https://doi.org/10.1098/rsos.210787}{doi:
		\nolinkurl{10.1098/rsos.210787}}.
	
	\bibitem{Moore2019}
	E.~J. Moore, S.~Sirisubtawee and S.~Koonprasert, \textit{A
		{C}aputo--{F}abrizio fractional differential equation model for {HIV}/{AIDS}
		with treatment compartment}, Adv. Differ. Equ. \textbf{2019} (2019), 200,
	\href{https://doi.org/10.1186/s13662-019-2138-9}{doi:
		\nolinkurl{10.1186/s13662-019-2138-9}}.
	
	\bibitem{MuthupandiGnanaprakasamBoulaaras2026}
	V.~Muthupandi, A.~J. Gnanaprakasam and S.~Boulaaras, \textit{Fractional order
		modeling of hepatitis {C} transmission dynamics with physics-informed neural
		network solutions}, BMC Infect. Dis. \textbf{26} (2026), 457,
	\href{https://doi.org/10.1186/s12879-026-12792-y}{doi:
		\nolinkurl{10.1186/s12879-026-12792-y}}.
	
	\bibitem{NowakMay2000}
	M.~A. Nowak and R.~M. May, \textit{Virus Dynamics: Mathematical Principles of
		Immunology and Virology}, Oxford University Press, 2000,
	\href{https://doi.org/10.1093/oso/9780198504184.001.0001}{doi:
		\nolinkurl{10.1093/oso/9780198504184.001.0001}}.
	
	\bibitem{PerelsonNelson1999}
	A.~S. Perelson and P.~W. Nelson, \textit{Mathematical analysis of {HIV}-1
		dynamics in vivo}, SIAM Rev. \textbf{41} (1999), 3--44,
	\href{https://doi.org/10.1137/S0036144598335107}{doi:
		\nolinkurl{10.1137/S0036144598335107}}.
	
	\bibitem{Perko2001}
	L.~Perko, \textit{Differential Equations and Dynamical Systems}, Springer,
	2001, 3rd ed.,
	\href{https://doi.org/10.1007/978-1-4613-0003-8}{doi:
		\nolinkurl{10.1007/978-1-4613-0003-8}}.
	
	\bibitem{Podlubny1999}
	I.~Podlubny, \textit{Fractional Differential Equations}, Academic Press, 1999.
	
	\bibitem{Selesi2024VirusDiffusion}
	D.~Sele{\v{s}}i, \textit{Virus diffusion modeling via fractional stochastic
		differential equations}, in: M.~Chatzakou, M.~Ruzhansky and D.~Stoeva (Eds.),
	\textit{Women in Analysis and {PDE}}, vol.~5 of \textit{Trends in
		Mathematics}, Birkh{\"a}user, Cham, 2024, pp. 333--342,
	\href{https://doi.org/10.1007/978-3-031-57005-6_34}{doi:
		\nolinkurl{10.1007/978-3-031-57005-6_34}}.
	
	\bibitem{SelesiTosic2025}
	D.~Sele{\v{s}}i and S.~To{\v{s}}i{\'c}, \textit{A fractional stochastic model
		for aerosol transmission of fluid droplets and virus exposure in closed
		spaces}, Filomat \textbf{39} (2025), 7647--7681,
	\href{https://doi.org/10.2298/FIL2522647S}{doi:
		\nolinkurl{10.2298/FIL2522647S}}.
	
	\bibitem{VanDenDriesscheWatmough2002}
	P.~van den Driessche and J.~Watmough, \textit{Reproduction numbers and
		sub-threshold endemic equilibria for compartmental models of disease
		transmission}, Math. Biosci. \textbf{180} (2002), 29--48,
	\href{https://doi.org/10.1016/S0025-5564(02)00108-6}{doi:
		\nolinkurl{10.1016/S0025-5564(02)00108-6}}.
	
	\bibitem{Wacker2024}
	B.~Wacker et~al., \textit{Revisiting the classical target cell limited
		dynamical within-host {HIV} model}, Math. Biosci. Eng. \textbf{21} (2024),
	5921--5948,
	\href{https://doi.org/10.3934/mbe.2024343}{doi:
		\nolinkurl{10.3934/mbe.2024343}}.
	
	\bibitem{Wodarz2007}
	D.~Wodarz, \textit{Killer Cell Dynamics: Mathematical and Computational
		Approaches to Immunology}, Springer, New York, 2007,
	\href{https://doi.org/10.1007/978-0-387-68733-9}{doi:
		\nolinkurl{10.1007/978-0-387-68733-9}}.
	
	\bibitem{Yamaguchi2024}
	D.~Yamaguchi, R.~Shimizu and R.~Kubota, \textit{Development of a {SARS-CoV-2}
		viral dynamic model for patients with {COVID-19} based on the amount of viral
		{RNA} and viral titer}, CPT: Pharmacomet. Syst. Pharmacol. \textbf{13}
	(2024), 1354--1365,
	\href{https://doi.org/10.1002/psp4.13164}{doi:
		\nolinkurl{10.1002/psp4.13164}}.
	
	\bibitem{ZhangHuSun2018}
	S.~Zhang, L.~Hu and S.~Sun, \textit{The uniqueness of solution for initial
		value problems for fractional differential equation involving the
		{C}aputo--{F}abrizio derivative}, J. Nonlinear Sci. Appl. \textbf{11} (2018),
	428--436,
	\href{https://doi.org/10.22436/jnsa.011.03.11}{doi:
		\nolinkurl{10.22436/jnsa.011.03.11}}.
	
	\bibitem{ZitzmannKaderali2018}
	C.~Zitzmann and L.~Kaderali, \textit{Mathematical analysis of viral replication
		dynamics and antiviral treatment strategies: From basic models to age-based
		multi-scale modeling}, Front. Microbiol. \textbf{9} (2018), 1546,
	\href{https://doi.org/10.3389/fmicb.2018.01546}{doi:
		\nolinkurl{10.3389/fmicb.2018.01546}}.
	
	\bibitem{ZitzmannKeRibeiroPerelson2024}
	C.~Zitzmann, R.~Ke, R.~M. Ribeiro and A.~S. Perelson, \textit{How robust are
		estimates of key parameters in standard viral dynamic models?}, PLOS Comput.
	Biol. \textbf{20} (2024), e1011437,
	\href{https://doi.org/10.1371/journal.pcbi.1011437}{doi:
		\nolinkurl{10.1371/journal.pcbi.1011437}}.
	
\end{thebibliography}

\end{document}